\documentclass[11pt,draftclsnofoot,journal,onecolumn]{IEEEtran}

\usepackage{amsmath}
\usepackage{amssymb}
\usepackage[hyphens]{url}
\usepackage{color}
\usepackage{flushend}
\usepackage{graphicx}
\usepackage{eurosym}
\usepackage{threeparttable}
\usepackage{algorithm}
\usepackage{algpseudocode}
\usepackage[tight,footnotesize]{subfigure}
\usepackage{bmpsize}

\usepackage{epstopdf}

\newtheorem{proposition}{Proposition}
\newtheorem{corollary}{Corollary}

\algtext*{EndLoop}
\algtext*{EndFor}
\algtext*{EndIf}
\algtext*{EndWhile}
\algtext*{EndFunction}

\newcommand{\textsubscript}[1]{$_{\text{#1}}$}

\newcommand{\txStateOff}{\textsf{S\textsubscript{OFF}}\,}
\newcommand{\txStateOn}{\textsf{S\textsubscript{ON}}\,}
\newcommand{\txStateProbe}{\textsf{S\textsubscript{PROBE}}\,}
\newcommand{\rxStateIdle}{\textsf{S\textsubscript{IDLE}}\,}
\newcommand{\rxStateWaiting}{\textsf{S\textsubscript{WAIT}}\,}
\newcommand{\rxStateCharged}{\textsf{S\textsubscript{CHARGED}}\,}

\newcommand{\txTimeoutBeaconing}{$t^{\text{ETx}}_\text{CrgReq}$}
\newcommand{\txTimeoutPowerProbeResponse}{$t^{\text{ETx}}_\text{PwrProbeRsp}$}
\newcommand{\txTurnOffFirstPwrProbeNotReceivedAfter}{$t^\text{ETx}_\text{turnOff}$}
\newcommand{\txTimeoutProbeReceive}{$t^\text{ETx}_\text{PwrProbe}$}
\newcommand{\txRssiThreshold}{$\eta^\text{ETx}_\text{CommTh}$}
\newcommand{\txRandomWaitTimeMax}{$t^\text{ETx}_\text{RandWait}$}
\newcommand{\rxDelayPing}{$t^\text{ERx}_\text{Ping}$}
\newcommand{\rxDelayRemoveLastProbeSender}{$t^\text{ERx}_\text{RmvLast}$}
\newcommand{\rxDelayWaitingForPower}{$t^\text{ERx}_\text{WaitForPwr}$}
\newcommand{\rxDelayPowerProbing}{$t^\text{ERx}_\text{PwrProbe}$}
\newcommand{\rxVoltageThreshold}{$\eta^\text{ERx}_\text{PowerTh}$}

\newcommand{\chargingRequest}{\texttt{REQ\textsubscript{CRG}}\,}
\newcommand{\powerProbeReport}{\texttt{REP\textsubscript{PWR}}\,}
\newcommand{\powerProbeRequest}{\texttt{REQ\textsubscript{PWR}}\,}

\newcommand{\txAddressQueue}{Q\textsubscript{TX}\,}
\newcommand{\tSyn}{t\textsubscript{SYN}\,}

\newcommand{\subparagraph}{}

\algnewcommand{\LineComment}[1]{\State \(\triangleright\) #1}
\algnewcommand{\Trigger}[1]{{\footnotesize \Statex/*\,\texttt{#1}\,*/}}
\algrenewcommand\algorithmicfunction{\textbf{upon}}

\makeatletter
\newcommand{\newalgname}[1]{%
 \renewcommand{\ALG@name}{#1}%
}
\newalgname{\footnotesize{Protocol}}

\IEEEoverridecommandlockouts

\renewcommand{\IEEEQED}{\IEEEQEDopen}

\begin{document}

\title{Green Wireless Power Transfer Networks}
\author{Qingzhi Liu, Micha{\l} Goli\'{n}ski, Przemys{\l}aw Pawe{\l}czak, and Martijn Warnier
\thanks{Qingzhi Liu, Micha{\l} Goli\'{n}ski and Przemys{\l}aw Pawe{\l}czak are with the Department of Electrical Engineering, Mathematics and Computer Science, Delft University of Technology, Mekelweg 4, 2600 GA Delft, The Netherlands (email: \{q.liu-1, p.pawelczak\}@tudelft.nl, m.golinski-1@student.tudelft.nl).}
\thanks{Martijn Warnier is with the Department of Technology, Policy and Management, Jaffalaan 5, 2628 BX Delft, The Netherlands (email: m.e.warnier@tudelft.nl). Qingzhi Liu is also affiliated with this department.}
\thanks{This work has been supported by the SHINE project of Delft Institute for Research on ICT and by the Dutch Technology Foundation STW under contract 12491.}}

\maketitle

\begin{abstract}
A Wireless Power Transfer Network (WPTN) aims to support devices with cable-less energy on-demand. Unfortunately, wireless power transfer itself---especially through radio frequency radiation rectification ---is fairly inefficient due to decaying power with distance, antenna polarization, etc.. Consequently, idle charging needs to be minimized to reduce already large costs of providing energy to the receivers and at the same time reduce the carbon footprint of WPTNs. In turn, energy saving in a WPTN can be boosted by simply switching off the energy transmitter when the received energy is too weak for rectification. Therefore in this paper we propose, and experimentally evaluate, two ``green'' protocols for the control plane of static charger/mobile receiver WPTN aimed at optimizing the charger workflow to make WPTN green. Those protocols are: `beaconing', where receivers advertise their presence to WPTN, and `probing' exploiting the receiver feedback from WTPN on the level of received energy. We demonstrate that both protocols reduce the unnecessary WTPN uptime, however trading it for the reduced energy provision, compared to the base case of `WPTN charger always on'. For example, our system (in our experiments) saves at most $\approx$80\% of energy and increases 5.5 times the efficiency with only $\approx$17\% less energy possibly harvested.
\end{abstract}

\section{Introduction}
\label{sec:introduction}

\begin{figure}
\centering
\subfigure[Retrieved: 21 March 2015]{\includegraphics[width=0.43\columnwidth]{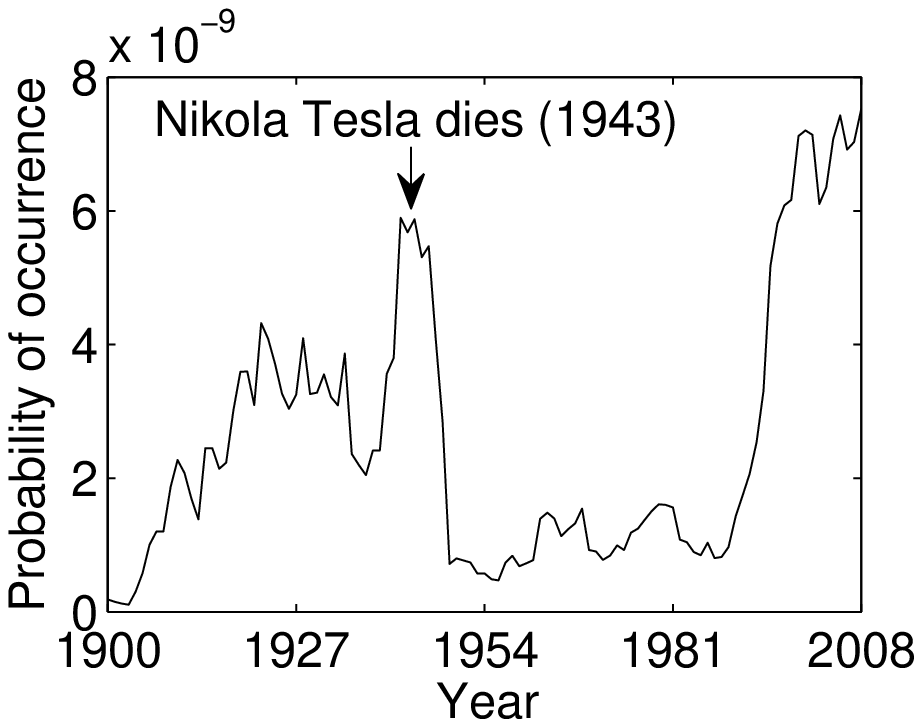}\label{fig:ngrams}}
\subfigure[Retrieved: 21 March 2015]{\includegraphics[width=0.43\columnwidth]{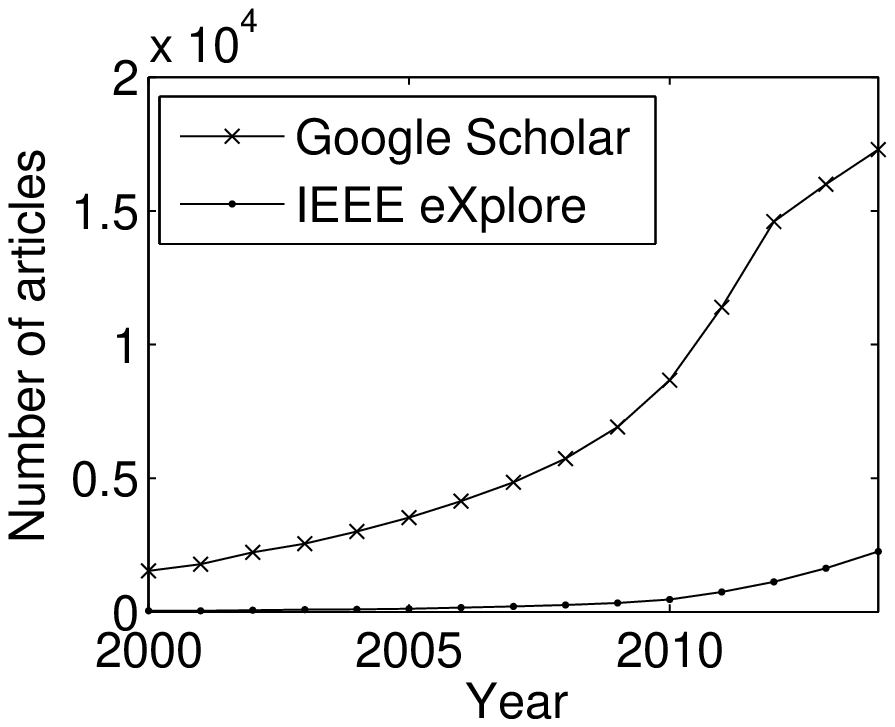}\label{fig:explore_scholar}}
\caption{Probbaility of `wireless power' $n$-gram occurrance extracted from Google digitalized books database~\cite{michael_science_2011} (left), and number of articles containing phrase `wireless power' published in a given year according to Google Scholar and IEEE eXplore database (right) (see also~\cite[Fig. 1]{massa_procieee_2013}).}
\label{fig:wireless_power_popularity}
\end{figure}

Edholm's law states that data rates offered by wire-less communication systems will converge with wired ones, where forward rate extrapolation indicates convergence around the year 2030~\cite{cherry_spectrum_2004}. As a result, the only cables that would require removal are the cables supporting power. In turn, wireless power (transfer) (WPT) is rapidly gaining momentum, see Fig.~\ref{fig:wireless_power_popularity}, and more companies are trying to capitalize on the wireless energy promise, refer for example to WiTricity~\cite{witricity_website}, uBeam~\cite{ubeam_website}, Ossia~\cite{ossia_website}, Artemis~\cite{artemis_website}, Energous~\cite{energous_website}, or Proxi~\cite{proxi_website}. 

A natural next step is the deployment of networks of WPT nodes (denoted throughout this work as WPTNs), i.e. deployed and dedicated WPT devices providing power to nearby energy receivers~\cite{dai_tpds_2014,xie_wcom_2013} (see also an example of a fully energy autonomous WPTN in~\cite[Fig. 1]{wicaksono_vtc_2011}). WPTNs are expected to find numerous applications, e.g. in sensing systems (vide rechargeable sensor network~\cite{he_tmc_2013}), biology research (vide insect fly monitoring~\cite{thomas_jbcs_2012} or animal inspection~\cite{greene_unpublished_2010}), or implantable networks (vide brain-machine interface~\cite{holleman_biocas_2008}). In all the above applications, the use of batteries is prohibitive (in biology-related applications---due to induced weight or prohibitive cabling, in implantable applications---due to necessity of surgical battery replacement), thus WPT is the only long-term viable option. Finally, we speculate that due to continuous decrease of energy consumption of embedded platforms~\cite[Fig. 1]{long_cicc_2008},~\cite[p. 87]{patel_wcm_2010}, within a decade energy provision through WPTNs will exceed the required energy cost of the above applications.

\subsection{Problem Statement}
\label{sec:problem_statement}

The necessity of WPT becomes immanent while observing that powering battery-based platforms using energy harvesting alone is not enough~\cite[Section I]{Pinuela_mtt_2013}. Simply put, ambient energy harvesting does not guarantee sufficient quality of energy provision~\cite[Sec. 3.3]{dai_tpds_2014}~\cite[Section I]{timotheou_twc_2014}. For example, large-scale London, UK-based RF far field energy harvesting measurements at 270 London Underground stations demonstrate that in the best case only 45\% of such locations can sustain the required minimum rectifying operation~\cite[Table VI]{Pinuela_mtt_2013} (for a single input source, considering digital TV, GSM 900/1800 and 3G transmitters).

However, WPT(N) has its own inherent deficiency. While there is a huge focus on making WPT(N) more efficient considering its hardware and physics, its energy conversion coefficient is still low~\cite[Section I]{Pinuela_mtt_2013} and absolutely not considered to be ``green''. This leads to a large carbon footprint of WPTN, which will be amplified as the technology spreads to the mass market. 

\subsubsection{Case Study---Cost of Uptime for State-of-the-Art WPT Nodes}
\label{sec:case_study_measurement}

\begin{table}
\centering
\caption{Energy Consumption of various WPTN devices, see Section~\ref{sec:problem_statement}}
\label{table:wptn_energy_consumption}
\begin{threeparttable}
\scriptsize
\begin{tabular}{| r | r | r | r | r |}
\hline
~ & R1000 & S420 & PWC & TI \\
\hline\hline
Consumed power (idle state) [W] & 24.31 & 5.57 & 4.13 & $\approx$0\tnote{a}\\
\hline
Consumed power (charge state) [W] & 45.20 & 12.68 & 4.13 & 1.59\\
\hline
\end{tabular}
\begin{tablenotes}
\scriptsize \item R1000---Impinj Speedway R1000 RFID reader~\cite{r1000_data_sheet} (firmware: Octane 3.2.4.240, hardware revision: 030-02-00001); S420---Impinj Revolution R420 RFID reader~\cite{s420_data_sheet} (firmware: 4.8.3.240, hardware version: 250-004-000); PWC---Powercast TX91501-3W transmitter~\cite[/products/powercaster-transmitters]{powercast_website}; TI---Texas Instruments BQ500410AEVM-085 transmitter evaluation board~\cite[/product/bq500410a]{ti_website}. Both RFID readers were controlled by~\cite{sllrp_github}, charge state induced in the inventory state of EPC Gen2 protocol.
\item[a] Value was too small to be measured by Energino.
\end{tablenotes}
\end{threeparttable}
\end{table}

From the monetary perspective, taking the energy conversion efficiency into account, assuming that the energy conversion coefficient at a given distance for wired and wireless system are $\eta_{\text{CPT}}=0.59$ and $\eta_{\text{WPT}}=0.01$, respectively, an extra cost of providing an equal amount of energy by WPT-based compared to a conventional cable-based energy supply during a period of 1\,year for a device consuming $\Gamma=12.5$\,Wh/day (approximate smartphone daily energy consumption) for a cost of $C_e=0.23$\,\euro/kWh is $365C_e\Gamma\eta_{\text{CPT}}/\eta_{\text{WPT}}=103.88$\,\euro. Therefore, it is of paramount importance to keep the chargers active only when the nearby receivers are requesting energy, and to forbid energy provision when charging (rectification) becomes ineffective\footnote{Finally, it is a truism to note that control of charging uptime minimizes unnecessary RF exposure~\cite[Sec. IX-H]{lu_arxiv_2014}.}.

To provide concrete results on how much power can be wasted we measured the power consumption of several WPT nodes using the Energino platform~\cite{gomez_wiopt_2012}---a realtime power consumption monitoring device for DC-powered devices. In the experiment, the Energino is connected between the mains power source and a WPT source to measure the power consumed by the charger. The set of example WPTN chargers is provided in Table~\ref{table:wptn_energy_consumption} representing (i) induction-based (short range) WPTN, i.e. TI board, (ii) RFID readers for transient computing/energy harvesting platforms~\cite{gollakota_computer_2014}, i.e. R1000 and S420, and (iii) long-range RF power transfer, i.e. PWC. We observe that all of the above devices consume non-negligible amount of energy, both RFID readers in particular, even in the idle state (except for the TI board).

\subsubsection{Research Question---How to make WPTN ``Green''?}
\label{sec:research_question}

In the most obvious WPTN topology (which naturally resembles cellular communication networks, where recent work proposed to overlay a WPTN on top of a cellular one~\cite{huang_arxiv_2012}) chargers are static but the energy receivers are moving and chargers and receivers are able to communicate with each other. Thus for energy receiver discovery and control of charger uptime, there needs to be a well designed control plane (and communication protocol) that turns on chargers only when the charging conditions are favorable---thus ``greenifying'' the WTPN.

Unfortunately, to the best of our knowledge, the problem of provision of energy to mobile energy receivers in such WPTN topology, guaranteeing fast charger discovery without the unnecessary energy waste at the charger has been overlooked. We conjecture that solving such problem is not a trivial task.

\subsection{Our Contribution}
\label{sec:contribution}

\begin{enumerate}
\item In this paper we prove that making WPTN ``green'' is \emph{an NP-hard problem}. In layman terms, we show that it is difficult to maximize harvested energy at the receiver and minimize idle time of the energy transmitters at the same time;

\item We then \emph{propose two heuristics, called Beaconing and Probing, that control the WPTN charge uptime} aiming at (i) maximization of harvested energy, charge accuracy, charge efficiency and (ii) minimization of energy consumed by the communication between energy receivers and energy transmitters and by the chargers;
 
\item Finally, \emph{we build (to the best of our knowledge) worlds-first green WPTN}. In our experiments, compared to a baseline case (all chargers being constantly on), system saves at most $\approx$80\% of energy and increases 5.5 times the efficiency with only $\approx$17\% less energy possibly harvested.

\end{enumerate}

\subsection{Paper Organization}
\label{sec:paper_organization}

The rest of the paper is organized as follows. Related work is discussed in Section~\ref{sec:related_work}. The WPTN model considered in this paper is introduced in Section~\ref{sec:wptn_model}. The formal problem statement of transmitter energy-conserving WPTN design is given in Section~\ref{sec:form_problem_statement}. Two proposed WPTN ``Green'' charge control protocols are briefly introduced in Section~\ref{sec:charger_energy_saving_protocol}, with their implementation details (and their performance evaluation) given in Section~\ref{sec:experimental_verification}. Experimental results are presented in Section~\ref{sec:experiment_result}. Limitations of this work and future challenges are presented in Section~\ref{sec:discussion}. Finally, the paper is concluded in Section~\ref{sec:conclusions}.

\section{Related Work: WPTN Chargers Uptime Control}
\label{sec:related_work}

A plethora of papers considers an information theoretic, or `classical' communications approach to analyze WPTNs, e.g. through Shannon capacity formulation of energy transfer, see e.g.~\cite{ju_twc_2014}, or optimization of transmission parameters of WPT sources to maximize considered objectives such as (i) harvested power~\cite{krikidis_tcom_2013,yang_arxiv_2013}, (ii) interference to collocated transmission sources~\cite{timotheou_twc_2014}, (iii) energy outage~\cite{huang_arxiv_2012,ng_globecom_2014}, (iv) charging delay~\cite{fu_infocom_2013} and (v) quality of service~\cite{liu_net_2014}. In the majority of those studies a continuous energy source is assumed, i.e. energy transmitters (ETx(s)) are always on/up, despite of energy receivers (ERx(s)) absence in the vicinity~\cite{timotheou_twc_2014,liu_net_2014} or are triggered at predefined intervals~\cite[Sec III-A]{mercier_jssc_2011}. Minimizing energy consumption while maximizing energy supply mostly limits itself to power control of ETx.

On-demand energy provision has been considered in~\cite{wang_tmc_2014} (in case of mobile ETx and static ERx),~\cite{wicaksono_vtc_2011,naderi_twc_2014} (for static ETx/ERx). In all of these works no actual protocol for controlling ETx uptime has been introduced. Papers that do propose ETx uptime control are~\cite{xiang_pimrc_2013} (controlling power flow in an inductive-based WPT), \cite{yoon_ccnc_2013} (although considered only architecturally without further investigation),~\cite[Fig. 4]{wicaksono_vtc_2011} (without any discussion on the details of the protocol), or~\cite{dai_infocom_2014} (in the context of electromagnetic exposure minimization). The most relevant work~\cite{naderi_twc_2014}, proposes a new medium access control protocol for WPTN-enabled sensor networks, which controls (among other things) ETx uptime, (i) considers both static ETx/ERx, and (ii) is ETx-centric, i.e. receivers must take care of requesting for energy (ETx never offer to send energy). Another relevant protocol has been considered in~\cite[Fig. 4]{wicaksono_vtc_2011}, but without analysis of the protocol parameters and its influence on the WTPN performance.

It is important to state that in all above works the number of ETx and ERx is always constant and an ERx/ETx discovery mechanism has been overlooked. Furthermore, as none of the above works (except for~\cite{naderi_twc_2014}) propose an actual charge control protocol for WPTNs, it is unknown how control and signaling affects provisioned energy for topologies other than that of~\cite{naderi_twc_2014}. We thus conclude that ETx charge control with discovery mechanism, saving energy due to signaling, has not been considered. This paper will fill this gap.

\section{WPTN Model}
\label{sec:wptn_model}

We now present the classification of the existing WPTNs. Based on this classification we shall select a WPTN model.

\subsection{WPT Classification}
\label{sec:wpt_classification}

\subsubsection{WPT Physical Layer Techniques}

The obvious classification in WPT relates to the source of energy which is later converted to electric current---please refer to recent surveys of WPT considering far field~\cite{visser_procieee_2013} (through various radio frequency (RF) ranges),~\cite{strassner_procieee_2013} (through microwave RF), and through inductive coupling~\cite{sample_procieee_2013} (near field),~\cite{ho_procieee_2013} (mid-field). The majority of WPTN that we are aware of are RF conversion-based---refer also to a recent survey of~\cite{lu_arxiv_2014,lu:2014:arxiv} considering design issues in RF rectification conversion for wireless networks.

\subsubsection{WPTN Topology}

A WPNT topology is composed of $m$ ETx and $n$ ERx. Consequently, four special cases WPNT are observed in the literature: with (i) $m=1$, $n>1$ e.g.~\cite{ju_twc_2014,zhao_tmc_2014}, (ii) $m>1$, $n>1$ e.g.~\cite{timotheou_twc_2014,naderi_twc_2014}, (iii) $m=1$, $n=1$ e.g.~\cite{liu_net_2014,yang_arxiv_2013}, and (iv) $m>1$, $n=1$ (which to the best of our knowledge has not been considered so far).

Considering mobility, a WPTN topology is categorized into: (i) static ETx/static ERx~\cite{liu_net_2014,naderi_twc_2014}, (ii) mobile ETx/static ERx~\cite{zhao_tmc_2014,wang_tmc_2014}, (iii) static ETx/mobile ERx~\cite{dai_tpds_2014,he_tmc_2013,mercier_jssc_2011}, and (iv) mobile ETx/mobile ERx (which also has not been considered so far in the literature to the best of our knowledge). A related categorization on WPTN mobility can be found in~\cite[Table IX]{lu_arxiv_2014} considering routing algorithms in energy harvesting sensor networks. In addition, WPTN topologies can be categorized into (i) planned, e.g.~\cite{timotheou_twc_2014,ju_twc_2014,zhao_tmc_2014,liu_net_2014} and (ii) unplanned, e.g.~\cite{krikidis_tcom_2013}.

\subsubsection{WTPN Energy/Communication Separation}

Separation of energy provision and communication/control in WPTN can be categorized into (i) joint energy and information transmission (through power splitting)~\cite{timotheou_twc_2014,liu_tcom_2013,huang_arxiv_2012}, (ii) time division approach~\cite{ju_twc_2014,liu_net_2014}, and (iii) frequency division~\cite{xiang_pimrc_2013,wang_tmc_2014} (often in relation to inductive-based WPT). For a in-depth survey we again refer to~\cite[Sec. III-E]{lu_arxiv_2014} and~\cite{bi:2014:arxiv}.

\subsection{Selected WPT Technology}

As we show above, due to the large design space it is prohibitive to consider all WPTN topologies in one work. Therefore, we constrain ourselves to the following WPTN model, due to simplicity and resemblance to a cellular topology (see Section~\ref{sec:research_question})---forming a baseline for further studies.

\subsubsection{WPTN Nodes}

We utilize RF-based energy transfer, as it is the (i) least invasive, (ii) and its hardware is the smallest of all WPT techniques (allowing for implantation in biological organisms while keeping the charging distance long range~\cite{holleman_biocas_2008}).

\paragraph{ETx}
\label{sec:etx}

ETxs are assumed to be static, with their locations planned such as to guarantee a minimum needed energy supply at any place in space (which nevertheless does not preclude energy being below the rectification threshold for any ERx at any point of space-time).

No central controller coordinating a set of ETx is considered (in contrary to~\cite[Sec. III]{chiu_apnoms_2012}). Charging is performed at a frequency $f_p$\,MHz, e.g. $f_p=915$\,Mhz in case of Powercast TX91501-3W transmitter~\cite[/products/powercaster-transmitters]{powercast_website}. ETx do not posses MIMO capabilities, beam steering nor transmission power control, making an ETx (and whole WPTN) design simple.

\subsubsection{ERx}
\label{sec:erx}

\begin{figure}
\centering
\subfigure[ETx]{\includegraphics[width=0.43\columnwidth]{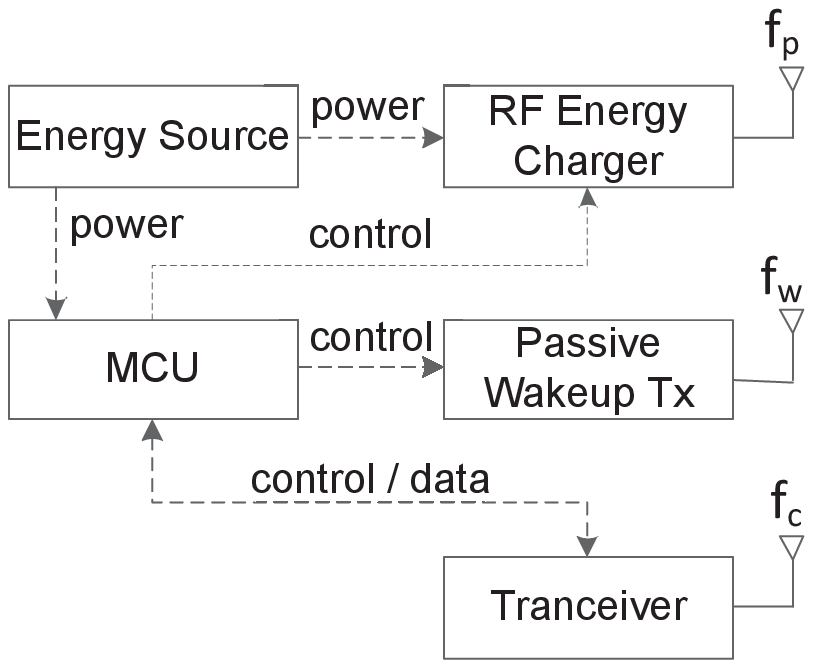}
\label{fig:wptn_hardware_model_charger}}
\subfigure[ERx]{\includegraphics[width=0.43\columnwidth]{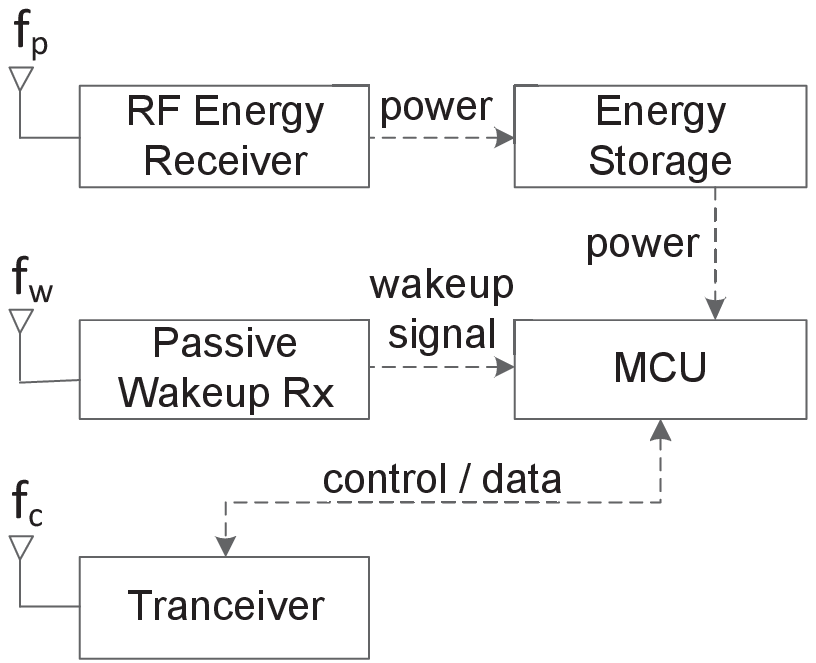}
\label{fig:wptn_hardware_model_receiver}}
\caption{WPTN components: ETx (left) and ERx (right).}
\label{fig:wptn_hardware_model}
\end{figure}

Charge requests/charge control between ETx and ERx is performed at frequency $f_c$, e.g. $f_c=2.4$\,GHz (as used in the experimental WPTN measurement setup introduced in Section~\ref{sec:experimental_verification} using XBee motes~\cite{xbee_website}). ERx aims at charging its internal battery/capacitor to the maximum level. ERx are mobile and equipped with wake-up radio capability, as in~\cite[Sec. II]{holleman_biocas_2008}, operating at frequency $f_w$\,MHz, e.g. $f_w=915$\,MHz~\cite{wisp_website}. Wake-up radio allows to conserve energy by ERx by avoiding idle listening to information broadcasted by ETx. In our WPTN charge protocol implementation we assume $f_w\neq f_c\neq f_p$ to avoid any interference scenarios (which does not preclude to design a WPTN charge control system with overlapping charge/wake-up/control frequencies). Note that ETx and ERx are schematically depicted in Fig.~\ref{fig:wptn_hardware_model}.

\subsubsection{WPTN Charge Control Protocol}
\label{sec:charge_request_protocol}

In general, as ERxs roam they are assumed to request a continuous flow of energy from the neighboring ETx. To achieve the ``green'' WPNT presented in Section~\ref{sec:problem_statement}, ETxs will send power to an ERx only when (i) a formal connection at frequency $f_c$ between ETx and ERx has been established and (ii) when the rectified energy at ERx is above the predefined threshold. Two attempts to introduce such protocol will be described in Section~\ref{sec:charger_energy_saving_protocol} and Section~\ref{sec:experimental_verification}. First, in the following section, we introduce the problem formally.

\section{Green WPTN---Formal Problem Statement}
\label{sec:form_problem_statement}

Let a WPTN be composed of $i \in \mathcal{M}, |\mathcal{M}| = m$ ERxs and $j \in \mathcal{N}, |\mathcal{N}| = n$ ETxs. The decision of ETx $j$ is to switch itself on or off (to conserve ETx power), with the switch decision denoted as $c_{j} \in \{1,0\}$, respectively. We assume that the decision $c_j$ is performed per time slot and that the state of WPTN (e.g. position of ETx and ERx, propagation conditions) is invariant within the time slot.

We consider the following WPTN performance parameters: (i) received power at ERx $i$ from ETx $j$ $\delta_{i,j}>0$ (expressed in Watts) noting that, as in~\cite[Sec. III-A]{dai_infocom_2014}, we ignore the effect of destructive interference~\cite[/PDF/P2110-datasheet.pdf (p. 7)]{ti_website}, i.e. $\sum_{\forall j: c_{j}=1}\delta_{i,j}$ increases with increasing $\sum_{j=1}^{n} c_{j}$, (ii) charging accuracy, which can be defined as\footnote{Note that $\vartheta _{i,j}\in[0,1]$ when described statistically over multiple time slots.}
\begin{equation}
\vartheta _{i,j}\triangleq
\begin{cases}
1, & c_{j}=1 \text{\,and\,} \delta_{i,j}\geq E_{r} \text{\,or\,} c_{j}=0 \text{\,and\,} \delta_{i,j}<E_{r},\\
0, & c_{j}=1 \text{\,and\,} \delta_{i,j}< E_{r} \text{\,or\,} c_{j}=0 \text{\,and\,} \delta_{i,j}\geq E_{r},
\end{cases}
\label{eq:accuracy_definition}
\end{equation}
where $E_{r}$ denotes minimum energy required for rectification, and (iii) charging efficiency
\begin{equation}
\xi _{i,j}\triangleq
\begin{cases}
\frac{\delta_{i,j}}{\mu_{j}}, & c_{j}\neq 1,\\
1, & c_{j} = 0,
\end{cases}
\label{eq:efficiency_definition}
\end{equation}
where $\mu_j$ (expressed in Watts) denotes the transmission power from ETx $j$, assuming that energy cost of running ETx $j$ is linearly proportional to $\mu_j$. It then follows that the charging error is $1 - \vartheta _{i,j}\triangleq\eta _{i,j}$ and energy wasting rate from ETx $j$ to ERx $i$ is $\frac{1}{\xi _{i,j}}\triangleq\psi _{i,j}$.

Furthermore, we introduce user defined WPTN performance constraints. To achieve safe charging $\sum\nolimits_{j = 1}^n \delta _{i,j}\leq \delta_t$ where $\delta_t$ is the exposure limit, see e.g.~\cite[Sec. III-B]{dai_infocom_2014}. To achieve WPTN-wide charging error $\sum\nolimits_{j = 1}^n \eta _{i,j}\leq \eta_t$, where $\eta_t$ is the acceptable error limit. Finally, to achieve WPTN-wide energy wasting rate $\sum\nolimits_{j = 1}^n \psi _{i,j}\leq \psi_t$, where $\psi_t$ is the allowed energy wasting limit.

In addition we introduce the following vectors: (i) $\mathbf{a}_{i,j} = [\delta_{i,j}, \vartheta_{i,j}, \xi_{i,j}]$ (vector of WPTN performance descriptors from ETx $j$ to ERx $i$), (ii) $\mathbf{b}_{i,j} = [\delta_{i,j} ,\eta_{i,j} ,\psi_{i,j}]$ (alternative form of $\mathbf{a}_{i,j}$), (iii) $\mathbf{s}_{t} = [\delta_t ,\eta_t ,\psi_t]$ (vector of WPTN-wide constraints). In addition we introduce $\mathbf{w}_{t}=[w_{\delta},w_{\eta},w_{\psi}]\in\mathbb{R}^{+}\cup\{0\}$ describing weights assigned to each WPTN performance descriptor.

We then define $o_{i,j}\triangleq\mathbf{w}_{t}\mathbf{a}_{i,j}^{T}$ (weighted sum of WPTN performance descriptors), $a_{i,j}\triangleq\mathbf{w}_{t}\mathbf{b}_{i,j}^{T}$ (weighted sum of alternative form of WPTN performance descriptors), and $s_{t}\triangleq\mathbf{w}_{t}\mathbf{s}_{t}^{T}$ (weighted sum of constraints). For $o_{i,j}$ we also define a total WPTN performance requirement $o_{q}$ (user specified). We can now introduce two problems formally
\begin{subequations}
\begin{align}
\text{\textbf{PI}:~}& \sum_{j = 1}^n \sum_{i = 1}^m o_{i,j}c_{j} \geq o_{q}, \label{eq:np_problem_1}\\	
\text{\textbf{PII}:~}\max & \sum_{j = 1}^n \sum_{i = 1}^m o_{i,j}c_{j}, \label{eq:np_problem_2}\\	
\text{\textbf{PI/PII} subject to~}& \sum_{j = 1}^n a_{i,j}c_{j} \leq s_{t}. \label{eq:np_problem_3}
\end{align}
\label{eq:np_problem_st}
\end{subequations}
\begin{proposition}
\textbf{PI} expressed as (\ref{eq:np_problem_1}) with subject to (\ref{eq:np_problem_3}) is NP-Complete. 
\begin{proof}
We will prove this proposition via restriction~\cite[Sec. 3.2.1]{Garey:1979:bell}. By allowing only instances of \textbf{PI} where $m=1$, $o_{i,j}=a_{i,j}$, $o_q=s_t= \frac{1}{2} \sum\nolimits_{j = 1}^n o_{i,j}$ and noting that $\sum\nolimits_{j=1}^{n}\sum\nolimits_{i=1}^{m}o_{i,j}c_{j}$ is a subset of all possible $o_{i,j}$ we restrict \textbf{PI} to the PARTITION problem which is NP-Complete~\cite[Sec. 3.1]{Garey:1979:bell}. Therefore \textbf{PI} is NP-Complete.
\end{proof}
\label{theo:np-complete}
\end{proposition}
\begin{corollary}
\textbf{PII} expressed as (\ref{eq:np_problem_2}) with subject to (\ref{eq:np_problem_3}) is NP-Hard. 
\begin{proof}
For $ \sum\nolimits_{j = 1}^n \sum\nolimits_{i = 1}^m o_{i,j}c_{j}\triangleq S_{\textbf{PI}} $ in \textbf{PI}, to decide whether $S_{\textbf{PI}} \geq o_q$, we ask \textbf{PII} to find $\max S_{\textbf{PI}} \triangleq S_{\textbf{PII}}$, and then check whether $S_{\textbf{PII}} \geq o_q$. If $S_{\textbf{PII}} \geq o_q$, then $S_{\textbf{PI}}\in[o_q,S_{\textbf{PII}}] $. However, if $S_{\textbf{PII}} < o_q$, then there is no $S_{\textbf{PI}} \geq o_q$. Then we have $\text{\textbf{PI}} \leq_{p} \text{\textbf{PII}}$, where $Y\leq_{p} X$ denotes ``$Y$ is polynomial time reducible to $X$''~\cite[Sec. 8.1]{kleinberg:2005:algo_design}. From Proposition~\ref{theo:np-complete} \textbf{PI} is NP-Complete. Therefore Problem II is NP-Hard, following from its definition~\cite[pp. 80]{leeuwen:1990:handbook}.
\end{proof}
\label{theo:np-hard}
\end{corollary}
We remark that \textbf{PI} and \textbf{PII} is a generalized case of~\cite[(3)]{dai_infocom_2014}.

Less formally \textbf{PI} and \textbf{PII} can be looked at as the multi-dimensional 0--1 knapsack problem (MKP)~\cite{freville2004multidimensional}. That is, the number of ERx $m$ with constraints $s_t$ corresponds to the number of knapsack with capacities. The number of ETx $n$ corresponds to the number of items. Each ETx $j$ generates $a_{i,j}$ in ERx $i$ and corresponds to each item consuming resources in the knapsack. Each ETx $j$ yields $o_{i,j}$ profits in receiver $i$ and corresponds to each item yielding profit in a knapsacks. Then, each ETx decides to turn on or off by assigning the values of $c_j$ and corresponds to each item being selected or not. Now, (i) the goal of \textbf{PI} is to decide whether the profit yielded by the ETx, i.e. $\sum\nolimits_{j = 1}^n \sum\nolimits_{i = 1}^m o_{i,j}c_{j}$, can be larger or equal to $o_q$ while not exceeding constraint $s_t$ in each receiver---while the decision goal of MKP is to decide whether the profit of the selected items can be larger or equal to the requirement and not exceeding the resource capacity of each knapsack. Similarly, (ii) the goal of \textbf{PII} is to make ETx yielding maximum profit from $\sum\nolimits_{j = 1}^n \sum\nolimits_{i = 1}^m o_{i,j}c_{j}$ while not exceeding constraint $s_t$ in each ERx---while the optimization goal of MKP is to make selected items yield maximum profit and not exceeding the resource capacity of each knapsack.

\section{Green WPTN: Charge Control Protocol Proposals}
\label{sec:charger_energy_saving_protocol}

Polynomial approximation schemes are used to solve MKP~\cite[Sec. 3.1]{freville2004multidimensional}. Nevertheless, this does not help us designing an algorithm for maximizing $o_{i,j}$ in WPTN, as \textbf{PI}/\textbf{PII} are introduced for a very simple (per time slot) WPTN system, not considering other elements that increase complexity of the problem (and the problem formulation), e.g. the mobility of ERx, the communication rate between ERx and ETx, or path loss. This shows a need to design a protocol to control $\mathbf{a}_{i,j}$. Therefore, in this paper we propose two simple protocols (heuristics) to solve (\ref{eq:np_problem_st}) in a best effort way. The general high-level idea behind these is as follows.

\begin{itemize}
\item Protocol 1---\emph{Beaconing}: when an ERx is in need of energy it broadcasts charging request packets periodically. If a charging request is received by an ETx, it turns itself on in order to charge ERx. While ETx is turned on, it expects that charging requests will arrive correctly at regular intervals from ERx.

\item Protocol 2---\emph{Probing}: Extending Protocol 1, if ERx is in need for energy it measures harvested energy and reports it to ETx. Then, ETx decides if ERx should be charged based on the information regarding energy harvested at ERx. If ETx does not receive any charge requests after timeout, then it will switch itself off.
\end{itemize}

To asses protocol 1 and 2 we propose a simple benchmark.

\begin{itemize}
\item Benchmark---\emph{Freerun}: (i) there is no communication among ERx and ETx, i.e. no energy is consumed for communication, (ii) ERx can freely harvest energy from each ETx, and (iii) ETxs are up all the time.
\end{itemize}
In the subsequent sections we will describe and evaluate experimentally Protocol 1 and 2 in detail.

\section{Green WPTN: Charge Control Protocol Implementation}
\label{sec:experimental_verification}

\subsection{WPTN Hardware, Software and Measurement Platform}
\label{sec:wptn_network_hardware}

\begin{figure}
\centering
\subfigure[ETx]{\includegraphics[height=0.33\columnwidth]{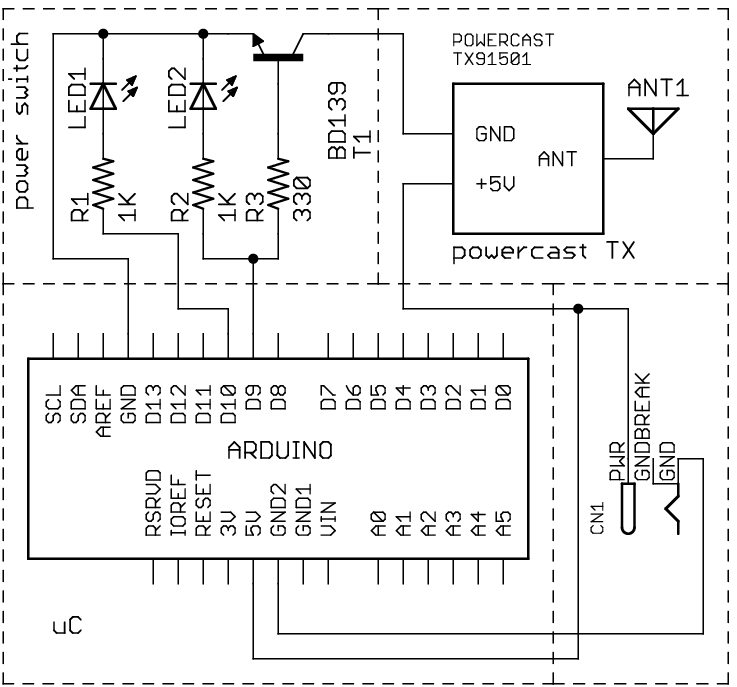}\label{fig:energy_transmitter}}
\subfigure[ERx emulator]{\includegraphics[height=0.33\columnwidth]{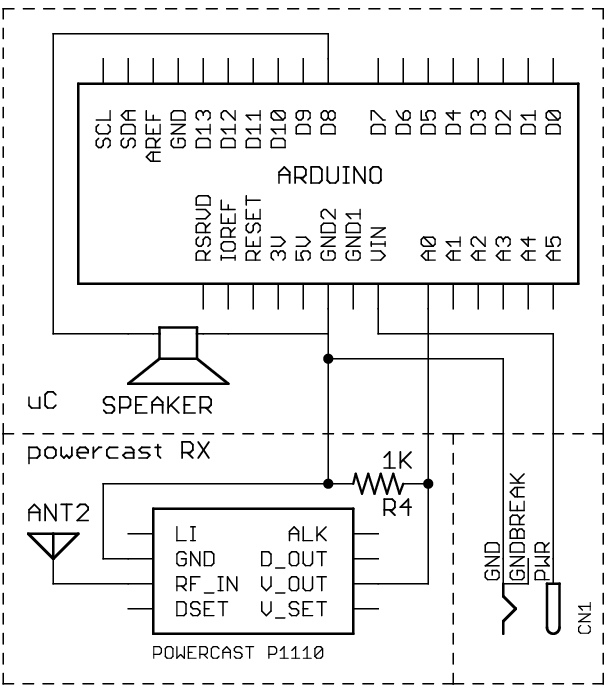}\label{fig:energy_receiver}}
\caption{Components of the implemented WPTN: (a) ETx, and (b) ERx emulator and a charge measuring unit. Notes: LED1 and LED2 are used for ETx state indication purpose; value of resistor R4 is user-changeable allowing to test the effect of various ERx impedances on the WPTN performance.}
\label{fig:wptn_hardware}
\end{figure}

To evaluate the proposed protocols we have deployed the following WPTN emulator, together with the charging protocol measurement platform. Our green WPTN is composed of four Powercast TX91501-3W transmitters with integrated antennas~\cite[/products/powercaster-transmitters]{powercast_website} and one P1110-EVB\footnote{Due to measurement simplicity we have chosen a harvester based on a continuous output power, P1110, rather than pulsed power, P2110.} receiver evaluation board with co-supplied 1\,dBi omnidirectional antenna~\cite[/products/development-kits]{powercast_website}, see Fig.~\ref{fig:wptn_hardware}. Each ETx, see Fig.~\ref{fig:energy_transmitter}, is connected with the mains power through the transistor switch controlled by the Arduino Uno board~\cite[/arduinoBoardUno]{arduino_website}. Analogically, ERx emulator is controlled by the same Arduino board, see Fig.~\ref{fig:energy_receiver}. 

All Arduino Uno boards are equipped with Wireless Secure Digital (SD) Shields~\cite[/ArduinoWirelessShield]{arduino_website} with Digi XBee IEEE 802.15.4 modules~\cite{xbee_website} attached (with 10EC version firmware and PCB antennas). XBee IEEE 802.15.4 modules are used to provide communication layer for the emulated WPTN. Each XBee device is configured to work as an end node in a frame-based API mode and given a unique 16-bit address. The rest of XBee IEEE 802.15.4 configuration parameters have default values. 

Each device logs its measurements and events to an SD card placed in the slot of the Wireless SD Shield. Both protocols introduced in Section~\ref{sec:charger_energy_saving_protocol}, as well as a measurement collection process, has been implemented in C++ amassing to more than 2800 lines of code.

There are two remarks that need to be made about our WPTN deployment. First, we note that we use the word `emulator' throughout, as ERx is still connected to the power supply. This was dictated by (i) the simplicity of the WPTN design, and (ii) an extra energy burden on ERxs due to data collection. Therefore, our WPTN implementation should be considered as an evaluation testbed for various WPTN protocols. Second, we note that the Powercast ETx/ERx we have used operated in the 915\,MHz center frequency channels (ISM Region 2), co-interfering with the Dutch KPN cellular operator and channels allocated to the Dutch Ministry of Defense.

\subsection{Green WPTN: Charge Control Protocol Details and Implementation}
\label{sec:wptn_network_protocol_implementation_experiment}

\subsubsection{Protocol Descriptors}

Before describing the operation of the two protocols in detail we introduce a set of support variables used by both protocols---messages and states---controlled by the timers provided in Table~\ref{table:implementation_parameters}.

\paragraph{ERx/ETx Messages}
\label{sec:p1_msgs}

Each packet from/to ERx/ETx is enclosed in an IEEE 802.15.4 frame, with frame header encapsulating source and destination address. In the protocols implementation of WTPN, on the reception of the packet, we allow to read received signal strength indicator of this particular packet. The following packet types used in our WPTN implementation are introduced:

\begin{itemize}

\item \chargingRequest: packet with charging request, broadcasted every \rxDelayPing\,s in Charging Request Phase (see Section~\ref{subsubsec:probing_based_protocol}) by ERx;

\item \powerProbeRequest: power report packet request sent by ETx from ERx used in Power Probing Phase (see Section~\ref{subsubsec:probing_based_protocol});

\item \powerProbeReport: packet containing two values: (i) voltage level on the load of ERx---$V$ and (ii) threshold level of ERx---\rxVoltageThreshold\footnote{The reason for sending $V$ and \rxVoltageThreshold from ERx to ETxs is due to ease of experiment result collection ($V$) and debugging (\rxVoltageThreshold). ETx uses \rxVoltageThreshold extracted from the packet instead of a pre-programmed one, therefore only ERx needs to be re-programmed in order to change this parameter of the experiment.}. \powerProbeReport can be a response to \powerProbeRequest (if in Power Probing Phase) or sent by ERx unsolicited (if in Charging Phase, see again Section~\ref{subsubsec:probing_based_protocol}).

\end{itemize}

\paragraph{ERx/ETx States}
\label{sec:messages}

ETx and ERx states are as follows.

\subparagraph{\textbf{ERx states:}}\label{sec:p1_erx_states} The following states are defined at ERx:

\begin{itemize}
\item \rxStateIdle: ERx is in need of energy and broadcasts \chargingRequest every \rxDelayPing\,s. ERx is in this state in Charging Request Phase of the protocol;

\item \rxStateWaiting: ERx awaits for the first power transmission from ETx. If power transmission is successful, ERx will move to \rxStateCharged. Else, if ERx does not receive any power for \rxDelayWaitingForPower\,s ERx moves back to \rxStateIdle\footnote{\rxStateIdle is a state in which ERx is initially when protocol is in Charging Phase.};

\item \rxStateCharged: State in which ERx is being charged by a specific ETx. ERx knows the address of ETx (last one stored in \txAddressQueue queue) and sends \powerProbeReport to this ETx every \rxDelayPowerProbing\,s. ERx is in this state if protocol is in Charging Phase and ERx harvests energy above \rxVoltageThreshold\,V.
\end{itemize}

\subparagraph{\textbf{ETx states:}} \label{sec:p1_etx_states} The following states are defined at ETx:

\begin{itemize}
\item \txStateOff: ETx does not transmit power;
\item \txStateOn: ETx does transmit power;
\item \txStateProbe: ETx probes ERx for power before decision to move to \txStateOff\ or \txStateOn; ETx waits in this state for \powerProbeReport from ERx for maximum of \txTimeoutPowerProbeResponse\,s. ETx is in this state when protocol is in Power Probing Phase.
\end{itemize}

\begin{table*}
\centering
\caption{Protocol parameter values used in the WPTN experiment implementation}
\vspace{-0.25cm}
\begin{threeparttable}
\begin{center}
\scriptsize
\begin{tabular}{| l | c c | l | c}
\hline
Symbol & \multicolumn{2}{c|}{Type\tnote{a}} & Description & Set value\\
\hline\hline
\txTimeoutBeaconing & B & & In \txStateOn---feedback timer within which unsolicited \chargingRequest packets from ERx need to be received & 8\,s\\
\hline
\txRssiThreshold & B & P & Received signal strength value below which a packet from ERx is ignored when received\tnote{b} & --70\,dBm\\
\hline
\rxDelayPing & B & P & Time between two consecutive \chargingRequest packets being broadcasted by ERx (if in \rxStateIdle\! for Probing) & 4\,s\\
\hline
\txTimeoutPowerProbeResponse & & P & Time ETx waits for \powerProbeReport after sending \powerProbeRequest & 4\,s\\
\hline
\rxDelayRemoveLastProbeSender & & P & Time for which each ETx address is stored in \txAddressQueue queue & 30\,s\\
\hline
\txTurnOffFirstPwrProbeNotReceivedAfter & & P & In \txStateOn---waiting time for the first (unsolicited) \powerProbeReport (sent by ERx on transition from \rxStateWaiting to \rxStateCharged\!) & 2\,s\\
\hline
\txTimeoutProbeReceive & & P & In \txStateOn---feedback timer within which unsolicited \powerProbeReport packets from ERx need to be received & 8\,s\\
\hline
\rxDelayPowerProbing & & P & In \rxStateCharged---time interval between two \powerProbeReport packets sent by ERx to the ETx currently charging ERx & 4\,s\\
\hline
\txRandomWaitTimeMax & & P & Maximum time ETx waits before sending \powerProbeRequest after receiving \chargingRequest from ERx\tnote{c}& 0.5\,s\\
\hline
\rxDelayWaitingForPower & & P & Maximum time ERx will wait for power from ETx while being in state \rxStateWaiting & 4\,s\\
\hline
\rxVoltageThreshold & & P & Voltage threshold for a load being attached to the microcontroller\tnote{d} & 0.5\,V\\
\hline
\tSyn & n/a & n/a & Measurement synchronization interval\tnote{e} & 2\,s\\
\hline
\end{tabular}

\begin{tablenotes}
\item[a] Protocol type: B---Beaconing, P---probing
\item[b] Parameter used to simulate different levels of communication layer power transmission/coverage
\item[c] Timer used to avoid collisions at ERx when multiple ETx hear the same \chargingRequest and send \powerProbeRequest immediately
\item[d] If the voltage level is above this threshold the power level is considered to be sufficient to initiate charging
\item[e] Value chosen as to synchronization happen more often than any events WPTN, see~\cite[Sec. 5.2]{golinski_msc_2015} for detailed discussion
\end{tablenotes}
\end{center}
\label{table:implementation_parameters}
\end{threeparttable}
\end{table*}

\subsubsection{Beaconing Protocol Details}
\label{subsubsec:beaconing_based_protocol}

The details of the protocol implementation is provided in Protocol~\ref{alg:beaconing_implementation}. Set of parameters describing the implementation are given in Table~\ref{table:implementation_parameters}. As a worst case scenario, in the implementation we assume that ERx is constantly in need for charging.

\begin{algorithm}[t]
\begin{threeparttable}[b]
\footnotesize
\caption{Beaconing---ETx and ERx events}
\begin{algorithmic}[1]
\Function{eventTimeoutPing}{\null}
\Comment \textbf{ERx event}---Note (a)
\State \Call{broadcast}{\chargingRequest}
\EndFunction

\smallskip

\Function{eventReceive}{\chargingRequest}
\Comment \textbf{ETx event}
	\State \Call{turnOnPowerTransmission}{\null}
	\State $\text{STATE} \leftarrow \txStateOn$
\EndFunction

\smallskip

\Function{eventTimeoutChargingRequest}{\null}
\Comment \textbf{ETx event}---Note (b)
	\State \Call{turnOffPowerTransmission}{\null}
	\State $\text{STATE} \leftarrow \txStateOff$
\EndFunction
\end{algorithmic}
\label{alg:beaconing_implementation}

\begin{tablenotes}
	\item[(a)] Executed every \rxDelayPing\,s
	\item[(b)] Executed in \txStateOn if \chargingRequest was not received for more than \txTimeoutProbeReceive\,s
\end{tablenotes}
\end{threeparttable}

\end{algorithm}

\subsubsection{Probing Protocol Details}
\label{subsubsec:probing_based_protocol}

The protocol executes in three phases described below. As in the case of Beaconing protocol, it is assumed that ERx constantly requires charging.

\paragraph{Charging Request Phase} In this phase ERx, every \rxDelayPing\,s, broadcasts \chargingRequest. At any time one or more ETx can receive \chargingRequest and initiate Power Probing Phase.

\paragraph{Power Probing Phase} Here ETx tries to find out if ERx is already being charged by another ETx. After ETx received \chargingRequest it will wait random time distributed uniformly with a maximum \txRandomWaitTimeMax\,s (used as a simple collision avoidance scheme at ERx) and then send \powerProbeRequest to ERx from which \chargingRequest was received. After ERx receives first \powerProbeRequest it will ignore all subsequent \powerProbeRequest packets from other ETxs in current Power Probing Phase. In return ERx sends \powerProbeReport containing current level of harvested energy. After ERx sends \powerProbeReport in Power Probing phase, it will wait predefined time of \rxDelayWaitingForPower\,s for the power transfer from ETx after which (if no power was transferred) it concludes that power transfer from ETx was unsuccessful.

If \powerProbeReport received by ETx contains power level lower than a power threshold, \rxVoltageThreshold, it means ERx is not currently harvesting energy and requires charging. Subsequently ETx tries to charge ERx and Charging Phase starts. If power will not be received, ERx will go back to Charging Request phase. In a process called blacklisting, ERx saves address of ETx that was unsuccessful in Charging Phase in its internal queue, denoted as \txAddressQueue. All the addresses are kept in \txAddressQueue for \rxDelayRemoveLastProbeSender\,s. If protocol is in a Power Probing state ERx ignores all ETxs with addresses stored in \txAddressQueue. This is done to prevent ETx that was not successful to initiate Power Probing Phase with given ERx again before the network conditions change, e.g. ERx moves to another position.

\paragraph{Charging Phase} After ETx starts charging ERx, there is a possibility that ERx harvests energy that is above \rxVoltageThreshold\,V. If this is the case ERx will start sending unsolicited \powerProbeReport to the current ETx (as ERx keeps track of ETx devices that tried to charge it). If \powerProbeReport packets are received by ETx at least every \txTimeoutProbeReceive\,s, ETx will continue charging a given ERx. If ERx does not receive enough power, it will not send \powerProbeReport packet to ETx within specified time, which will result in stop of power transmission from ETx to ERx.

Pseudocode of the Probing protocol is described formally in Protocol~\ref{alg:erx_states} and in Protocol~\ref{alg:etx_states}, for ERx and ETx side, respectively. Again, Table~\ref{table:implementation_parameters} summarizes all parameters of the protocol and their assumed values in the experiment.

\begin{algorithm}[t]
\begin{threeparttable}[b]
\footnotesize
\caption{Probing---ERx events}
\begin{algorithmic} [1]

\Function{eventTimeoutPing}{\null}
\Comment{Note (b)}
	\State \Call{broadcast}{\chargingRequest}
\EndFunction

\smallskip

\Function{eventReceive}{\powerProbeRequest, ETx $i$}
\Comment{Note (a)}
	\If{STATE = \rxStateIdle \textbf{and} ETx $i$ address $\notin$ \txAddressQueue}
		\State address ETx $i\rightarrow$ \txAddressQueue
		\State \Call{send}{\powerProbeReport, ETx $i$}
		\State $\text{STATE} \leftarrow \rxStateWaiting$
	\EndIf
\EndFunction

\smallskip

\Function{eventTimeoutWaitingForPower}{\null}
\Comment {Note (c)}
	\State $\text{STATE} \leftarrow \rxStateIdle$
\EndFunction

\smallskip

\Function{eventTimeoutPowerProbing}{\null}
\Comment {Note (d)}
	\State \Call{send}{\powerProbeReport, ETx $i$=\txAddressQueue(first)}
\EndFunction

\smallskip

\Function{eventVoltageAboveThreshold}{\null}
\Comment {Note (e)}
	\State \txAddressQueue$\leftarrow$\Call{Enqueue}{ETx $i$ address}
	\State \Call{send}{\powerProbeReport, ETx $i$}
	\State $\text{STATE} \leftarrow \rxStateCharged$
\EndFunction

 \smallskip

\Function{eventVoltageBelowThreshold}{\null}
\Comment{Note (f)}
	\State $\text{STATE} \leftarrow \rxStateIdle$
\EndFunction

\smallskip

\Function{eventTimeoutRemoveOldestProbeSender}{\null}
\Comment{Note (g)}
	\State \txAddressQueue$\leftarrow$\Call{Dequeue}{\txAddressQueue(last)}
\EndFunction

\end{algorithmic}
\label{alg:erx_states}

\begin{tablenotes}
	\item[(a)] When \powerProbeRequest received from ETx $i$
	\item[(b)] Every \rxDelayPing\,s if ERx is in \rxStateIdle
	\item[(c)] When ERx in \rxStateWaiting and no power from ETx for more than \rxDelayWaitingForPower\,s
	\item[(d)] Every \rxDelayPowerProbing\,s if ERx is in \rxStateCharged
	\item[(e)] When ERx in \rxStateWaiting receives power and load voltage exceeds \rxVoltageThreshold\,V
	\item[(f)] When in \rxStateCharged and attached load voltage drops below \rxVoltageThreshold\,V
	\item[(g)] When oldest address in \txAddressQueue has been stored longer than \rxDelayRemoveLastProbeSender\,s
\end{tablenotes}
\end{threeparttable}

\end{algorithm}

\begin{algorithm}[t]
\begin{threeparttable}[b]
\footnotesize
\caption{Probing---ETx events}
\begin{algorithmic}[1]

\Function{evenReceive}{\chargingRequest, ERx $i$}
\Comment{Note (a)}
	\If{$\text{STATE} = \txStateOff$}
		\State \Call{waitRandom}{\txRandomWaitTimeMax} \Comment{Uniform distribution}
		\State \Call{send}{\powerProbeRequest, $i$}
		\State $\text{STATE} \leftarrow \txStateProbe$
	\EndIf
\EndFunction

\smallskip
 
\Function{evenReceive}{\powerProbeReport, ERx $i$, $V$, \rxVoltageThreshold}
\Comment{Note (b)}
	\If{$\text{STATE} = \txStateProbe$}
		\If{$V \geq$ \rxVoltageThreshold}
			 \State $\text{STATE} \leftarrow \txStateOff$
		\Else
			\State \Call{turnOnPowerTransmission}{\null}
			\State $\text{STATE} \leftarrow \txStateOn$
		\EndIf
	\EndIf
\EndFunction

\smallskip

\Function{eventTimeoutPowerProbeResponse}{\null}
\Comment{Note (c)}
	\State $\text{STATE} \leftarrow \txStateOff$
\EndFunction

\smallskip

\Function{eventTimeoutOnPowerProbe}{\null}
\Comment{Note (d)}
	\State \Call{turnOffPowerTransmission}{\null}
	\State $\text{STATE} \leftarrow \txStateOff$
\EndFunction

\end{algorithmic}
\label{alg:etx_states}

\begin{tablenotes}
	\item[(a)] When \chargingRequest received from ERx $i$
	\item[(b)] When \powerProbeReport received from ERx $i$ with $V$\,V and \rxVoltageThreshold\,V
	\item[(c)] When in \txStateProbe after sending \powerProbeRequest the \powerProbeReport from ERx not received for more than \txTimeoutPowerProbeResponse\,s
	\item[(d)] When in \txStateOn and \powerProbeReport from ERx not received for more than \txTimeoutProbeReceive\,s
\end{tablenotes}
\end{threeparttable}

\end{algorithm}

\subsection{Synchronization in WPTN}
\label{sec:synchronization}

For accurate collection of measurements a time synchronization is implemented as follows~\cite[Ch. 5]{golinski_msc_2015}. ERx broadcasts its timestamp every \tSyn\,s. On the reception each ETx takes this timestamp as its own. After the experiment, timestamps received from ERx are subtracted from local ETx time. The result is the sum of transmission, processing and the actual clock time drift. Therefore time drift is a difference between the time of the reception of \chargingRequest at ETx and time of ERx broadcasting it. As it is impossible to eliminate processing and transmission time from the measurement in a simple way all measurements were made with the assumption that these values are negligible compared with other events in WPTN.

\subsection{WPTN Deployment and Experiment Scenarios}
\label{sec:experiment_scenario}

\begin{figure}
\centering
\subfigure[Measurement setup]{\includegraphics[height=0.33\columnwidth]{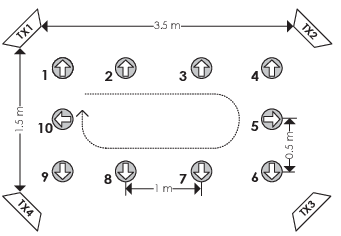}\label{fig:experiment_setup}}
\subfigure[ERx emulator/ETx pair]{\includegraphics[height=0.33\columnwidth]{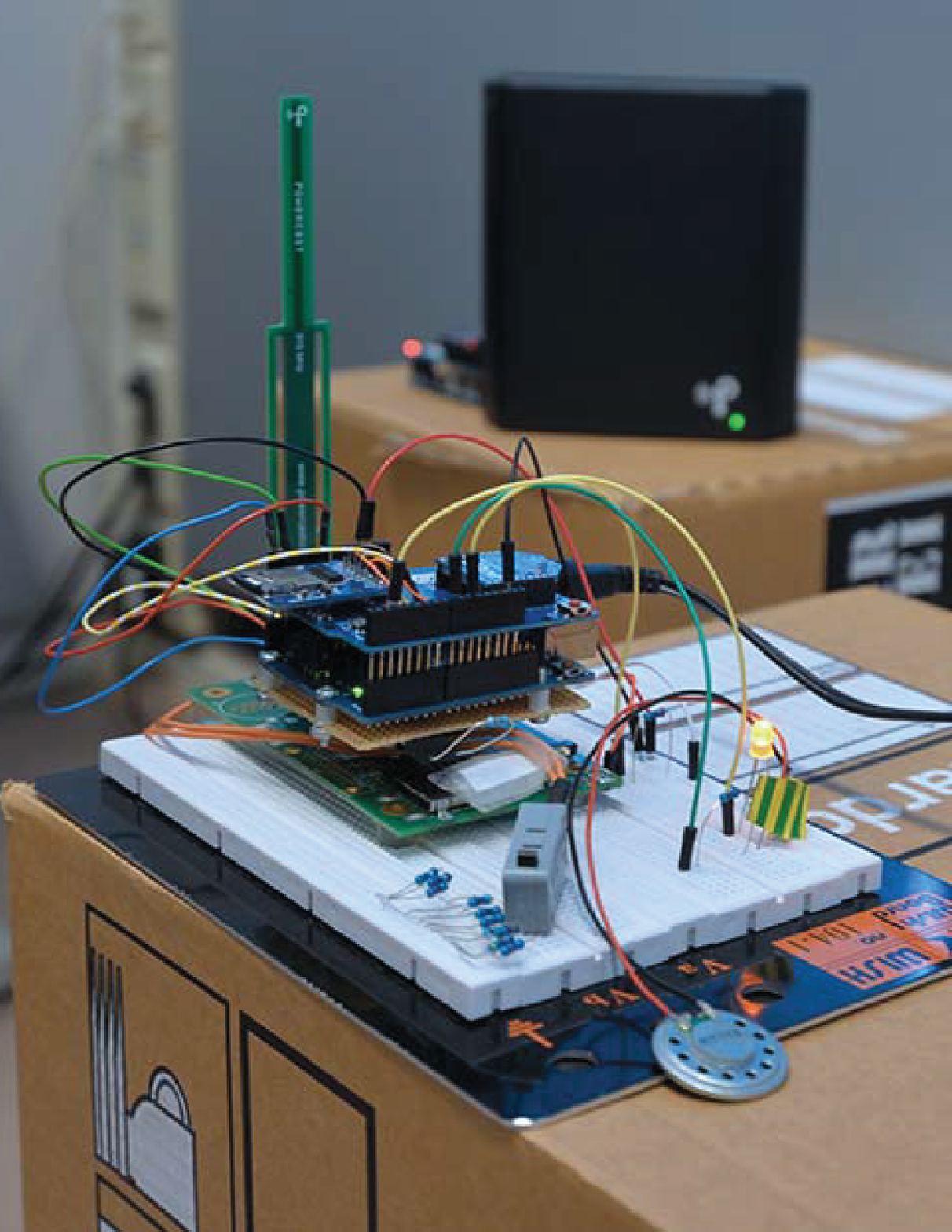}\label{fig:photograph_setup}}
\caption{WTPN experiment setup: (a) ERx---gray circles, with its position (marked as \emph{1--10}) and its orientation (marked with arrows), where dashed arrow denotes ERx movement direction; and ETx---white trapeziod, see also Section~\ref{sec:experiment_scenario}, (b) photograph of the ERx emulator/ETx pair in the laboratory setting---front: ERx emulator, back: one ETx.}
\label{fig:wptn_setup}
\end{figure}

All ETxs and an ERx emulator were placed on cardboard boxes 50\,cm tall---allowing for equal positioning in the vertical plane. Four ETxs were placed at the edges of a 1.5\,m\,$\times$\,3.5\,m rectangular plane. The angle of the front of the antennas were regulated and initially shifted 45 degrees to the border of the rectangular plane, with their center axis unchanged during the entire experiment. Conversely, the ERx emulator was allowed to be placed in ten different positions separated in vertical and horizontal axes by 1\,m and 0.5\,m, respectively. Front of the ERx emulator panel antenna was always vertical to the ground floor. Schematic representation of all ETxs and ERx emulator positions are presented in Fig.~\ref{fig:experiment_setup}. The measurement setup has been built inside the master student office of TU Delft Embedded Systems Lab, see Fig.~\ref{fig:photograph_setup}, with movement of humans during the experiment minimized.

Within such setup, the experiment simulated the random appearance/disappearance of the ERx in a controlled and replicable fashion. The experiment was started by placing the ERx emulator at position `1', see Fig.~\ref{fig:experiment_setup}, and initializing a measurement by turning on or pressing the reset button of each device in WPTN. From that moment the ERx emulator advertises itself to WPTN and starts collecting measurements. The ERx emulator is placed at this position for a random time chosen uniformly between 40\,s and 44\,s. This behavior is introduced to simulate random appearances and arrivals of the ERx emulator within one time period of sending \chargingRequest. After that time it signals the end of the single measurement through a buzzer, see Fig.~\ref{fig:energy_receiver}. Consequently, protocol execution is paused for 15\,s allowing the experiment operator to move ERx emulator to the next measuring position. One round of data collection is finished when the ERx emulator reaches position `10', with the movement pattern depicted in Fig.~\ref{fig:experiment_setup}. Each round of movements has been repeated five times for statistical significance. The duration of single experiment was ten minutes. Therefore, results presented in the following section are based on approximately nine hours of constantly running measurements. The voltage at resistor R4 of ERx, Fig.~\ref{fig:energy_receiver}, was sampled with a period of 0.1\,s.

\subsubsection{Experimental Results Replication}
\label{sec:experiment_result_replication}

For the results reproducibility measurement data, MATLAB post-processing scripts and Arduino-based charge control protocol implementation are available upon request or via~\url{http://www.es.ewi.tudelft.nl/reports/support_files/ES-2015-01_source_code.zip}.

\section{Experiment Results}
\label{sec:experiment_result}

\subsection{Performance Indicators of Green WPTN Control}
\label{sec:performance_descriptors}

We will look at the following performance indicators for both protocols.

\subsubsection{ERx Energy Harvested}
\label{sec:energy_harvested_metric}

Amount of energy harvested by ERx during the entire experiment.

\subsubsection{ETx Energy Consumption}
\label{sec:etx_energy_consumption}

Total energy consumed by all ETx during the entire experiment.

\subsubsection{WPTN Charging Efficiency}
\label{sec:}

Ratio of ERx energy harvested to the energy consumed by ETx during charging.

\subsubsection{ERx Energy Consumption}
\label{sec:energy_consumption_metric}

For a fair comparison of the two WPTN charge control protocols we take into consideration the energy consumed by the transmission/reception of packets from/to ERx. Avoiding extra burden of measuring the energy consumption of ERx communication (refer e.g. to~\cite{casilari_sensors_2010} for such studies) we directly calculated energy consumption values from the data sheet for the ERx emulator we built. The set of parameters used in the calculations are given in Table~\ref{table:hardware_variable_table}.

\begin{table}
\centering
\begin{threeparttable}
\caption{Variables used for ETx energy consumption calculation: ATmega328 (Arduino Uno's microcontroller) and Digi Xbee IEEE 802.15.4}
\label{table:hardware_variable_table}
\scriptsize
\begin{tabular}{| c | l | c |}
\hline
Symbol & Description & Value \\
\hline\hline
$U_{s}$ & Voltage supply: ATmega328/Xbee & 3.3\,V \\
$I_{T,x}$ & Current consumed: Xbee transmission & 35\,mA \\
$I_{R,x}$ & Current consumed: Xbee reception & 50\,mA \\
$I_{S,x}$ & Current consumed: Xbee sleep state & 10\,$\mu$A \\
$I_{S,a}$ & Current consumed: ATmega328 sleep state & 9\,$\mu$A \\
$I_{A,a}$ & Current consumed: ATmega328 active state & 1.7\,mA \\
$R_{d}$ & Digi Xbee data rate & 9.6\,kb/s \\
$S_{p}$ & Digi Xbee packet size & 960\,bits \\
\hline
\end{tabular}
\begin{tablenotes}
\item ATmega328 DC characteristics follow from~\cite[Table 29-7]{avr_data_sheet}
\item Digi Xbee DC characteristics follow from~\cite{xbee_data_sheet}
\end{tablenotes}
\end{threeparttable}
\end{table}

Although following Fig.~\ref{fig:wptn_hardware_model} we assume that the passive wakeup radio is used to wake up ERx from the off state to communication with ETx state, in the calculation we nevertheless include the cost of the idle state of the microcontroller and the radio of the ERx. Therefore, we calculate the ERx power consumption as $E_{c}=E_{c}^{(X)}+E_{c}^{(A)}$, where
\begin{align}
E_{c}^{(X)} & \approx E_{T_x}^{(X)} + E_{R_x}^{(X)} + E_{I}^{(X)} \nonumber\\
& = N_{t}\frac{S_p}{R_d}(U_{s}I_{T,x}) + N_{r}\frac{S_p}{R_d}(U_{s}I_{R,x})+T_{E}U_{s}I_{S,x},
\label{eq:communication_energy_x}
\end{align}
denoting total energy consumption of Digi Xbee board for $T_{E}$\,s total experiment time, composed of transmission energy ($E_{T_x}^{(X)}$), receive energy ($E_{R_x}^{(X)}$), and idle state energy ($E_{I}^{(X)}$), respectively, for and $N_{t}$ and $N_{r}$ are the number of packets that the ERx transmits and receives, and
\begin{align}
E_{c}^{(A)} & \approx E_{A}^{(A)} + E_{I}^{(A)} \nonumber\\
& = \left(N_{t}\frac{S_p}{R_d}+N_{r}\frac{S_p}{R_d}\right)U_{s}I_{A,a}+T_{E}U_{s}I_{S,a},
\label{eq:communication_energy_a}
\end{align}
is the energy consumed by the Arduino Uno board composed of active state energy ($E_{A}^{(A)}$), and idle state energy ($E_{I}^{(A)}$), respectively\footnote{Note that (\ref{eq:communication_energy_x}) and (\ref{eq:communication_energy_a}) are worst case approximations, we assume for simplicity during transmission and reception Arduino was simultaneously in sleep state---this is due to a small overhead of energy consumption by transmission and reception compared to the total time when the node was idle.}.

\newcommand{\tOptB}{T^{\text{B}}_{\text{opt}}}
\newcommand{\tPesB}{T^{\text{B}}_{\text{pes}}}
\newcommand{\tOptBavg}{\overline{T}^{\text{B}}_{\text{opt}}}
\newcommand{\tPesBavg}{\overline{T}^{\text{B}}_{\text{pes}}}
\newcommand{\tStartB}{T^{\text{B}}_{\text{start}}}
\newcommand{\tOptP}{T^{\text{P}}_{\text{opt}}}
\newcommand{\tPesP}{T^{\text{P}}_{\text{pes}}}
\newcommand{\tOptPavg}{\overline{T}^{\text{P}}_{\text{opt}}}
\newcommand{\tPesPavg}{\overline{T}^{\text{P}}_{\text{pes}}}
\newcommand{\tStartP}{T^{\text{P}}_{\text{start}}}

\subsubsection{Time to Charge}
\label{sec:time_to_charge_metric}

Finally, we measure and analytically evaluate protocol-specific parameter, i.e. time to charge---a time between transmission of charge request by the ERx to the beginning of charge provision by the first-responding ETx. 

We consider one ERx and $N$ ETxs, as in the experiment. ERx is in charging range of $K$ ETxs ($K \leq N$) and in communication range of all $N$ ETxs. At a given moment of time (given ERx position) $K$ and $N$ is fixed. Our goal is to derive formulas for expected time to charge an ERx in WPTN.

\paragraph{Beaconing}

In Beaconing implementation we assumed only one round of charging, after which all ETxs within the communication range of ERx will be turned on. The duration of this round is $\tOptB = \mathit{U}(0, t^\text{ERx}_\text{Ping})$, where $\mathit{U}(a, b)$ denotes an uniform distribution from $a$ to $b$.

If ERx randomly starts to send charge requests in WPTN, then time to charge is $\tStartB = \tOptB$. A cumulative distribution function (CDF) of $\tStartB$ under the assumption that $\tOptB$ is not random but constant and equal to its mean value, $\tOptBavg$, is
\begin{equation}
F_{\tStartB}(t) \approx t/t^\text{ERx}_\text{Ping},
\end{equation}
where $t\in[0,t^\text{ERx}_\text{Ping}]$.

\paragraph{Probing}

Probing works in rounds. Successful round starts with Charging Request, continues to Power Probing Phase and ends in Charging Phase, in which the protocol stays, successfully charging ERx. However, if the charging is not successful, the protocol goes back to the Charging Request phase. In the new Charging Request phase previous ETx, that unsuccessfully attempted to charge ERx, is excluded from WPTN. Therefore, for the first round, there are $N$ ETx and $K$ ETx that could charge ERx. If we choose one of $N-K$ ETxs that could not charge ERx, we exclude it in the next round, which starts with $N-1$ ETx and $K$ ETx that could charge ERx. Considering random variable $X_i$---ERx was charged in round $i$, then $\Pr(X_i)=K/N$ for $i=1$ and $\Pr(X_i) = \frac{K}{N-(i-1)} \prod_{j = 0}^{i-2}\frac{N-j-K}{N-j}$ for $i\in[2,N-K+1]$.

Length of a successful round, $\tOptP = \mathit{U}(0, t^\text{ERx}_\text{Ping})$, is different from the unsuccessful round, $\tPesP = \mathit{U}(0, t^\text{ERx}_\text{Ping}) + t^\text{ERx}_\text{WaitForPwr}$. Considering average values of those variables, $\tOptPavg$ and $\tPesPavg$, respectively, as a consequence the CDF of $\tStartP$ assumes no randomness of $\tOptP$ and $\tPesP$, giving
\begin{equation}
F_{\tStartP}(t) \approx \textstyle \sum_{i = 1}^{f(t)} \Pr(X_i), t \in \{0, (N-K) \tPesPavg + \tOptPavg\},
\end{equation}
where $f(t) = \left\lfloor \frac{ t - \tOptPavg}{ \tPesPavg} - 1 \right\rfloor$.

\subsubsection{WPTN Charge Accuracy}
\label{sec:charge_accuracy_metric}

\paragraph{Reference Measurement}

To calculate accuracy for both protocols we need to measure the reference case first. The reference will denote whether ETx should switch on during a particular time to charge ERx. We measure the reference scenario as follows.
\begin{enumerate}
\item We mark the appearance time, $t^{(a)}_{p}$, and disappearance time, $t^{(d)}_{p}$, of the ERx at each position depicted in Fig.~\ref{fig:experiment_setup}. The ERx stays at one location for 20\,s and is allowed to move to a new position within 15\,s from switch off, respectively\footnote{Note that these respective times were shorter than those during actual experiments, refer to Section~\ref{sec:experiment_scenario}.};
\item During each round of movement of the ERx, one ETx is charging at a time. After position 10 was reached by the ERx as given in Fig.~\ref{fig:experiment_setup}, a new ETx is turned on and a currently charging ETx is switched off;
\item We consider the following situation to be correct: if $V>$\rxVoltageThreshold at a resistor R4 of ERx, then ETx $j$ should switch on to charge the ERx at this position, otherwise it should switch off. Each event of voltage crossing threshold is added to a vector $\mathbf{R_c}\triangleq [t^{(a,1)}_{p}, t^{(d,1)}_{p}, t^{(a,2)}_{p}, t^{(d,2)}_{p},$ $\ldots,t^{(a,x)}_{p}, t^{(d,x)}_{p}]$, where $t^{(d,x)}_{p}<T_{E}$ and $t^{(x,y)}_{p}$, $x\in\{a,d\}$ denote the start ($x=a$) and stop ($x=d$) of the reference charge and $y\in \mathbb{N}$ denote its successive number. We note that the voltage sampling period at resistor R4 of ERx is 0.1\,s, just like in experiments in Section~\ref{sec:experiment_scenario}.
\end{enumerate}

\paragraph{Charge Accuracy Metric}

Having the reference case we can compare the actual working time sequence of each ETx $j$ (for each protocol---Beaconing and Probing) with the reference vector $R_c$ and calculate charge accuracy as\footnote{We refer the interested reader to~\cite[Ch. 5]{golinski_msc_2015} where other types of accuracy metrics (including ETx accuracy and ERx accuracy) are introduced.} 
\begin{equation}
\boxed{\overline \vartheta\triangleq (\mathbf{R_c}\Leftrightarrow \mathbf{R_c^{(x)}})/T_{E}},
\end{equation}
where $\mathbf{R_c^{(x)}}$ is the corresponding vector of for protocol $x$ and $\Leftrightarrow$ denotes XNOR operation.
\begin{figure}
\centering
\subfigure[ERx harvested energy]{\includegraphics[width=0.6\columnwidth]{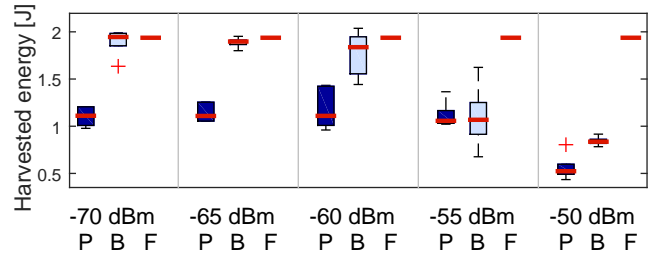}\label{fig:exp1_1}}
\subfigure[ETx consumed energy]{\includegraphics[width=0.6\columnwidth]{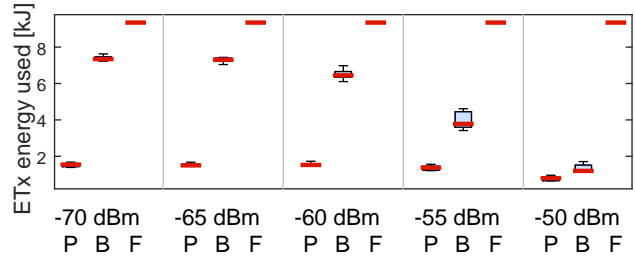}\label{fig:exp1_0}}
\subfigure[ERx communication cost]{\includegraphics[width=0.6\columnwidth]{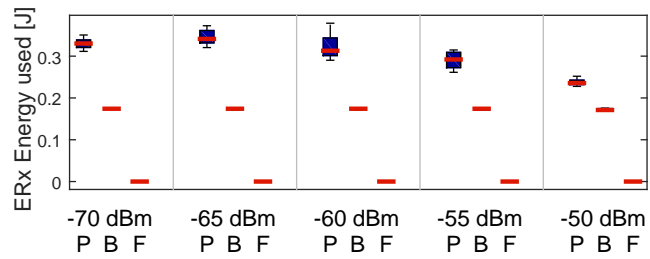}\label{fig:exp1_3}}
\subfigure[WPTN Efficiency]{\includegraphics[width=0.6\columnwidth]{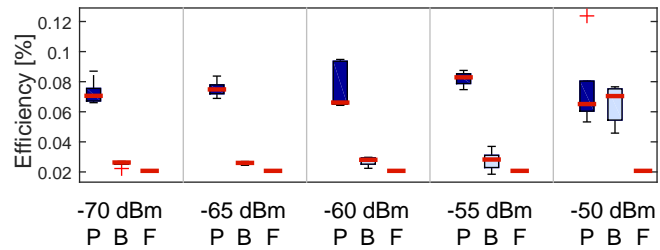}\label{fig:exp1_2}}
\subfigure[WPTN Accuracy]{\includegraphics[width=0.6\columnwidth]{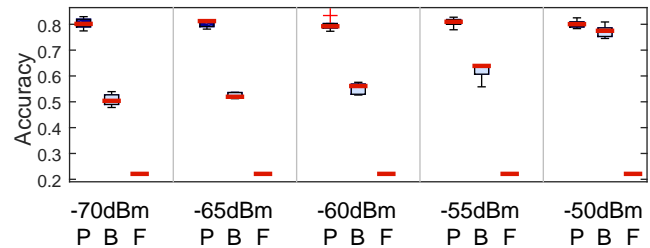}\label{fig:exp1_4}}
\caption{WTPN experiment results: P---Probing, B---Beaconing, F---Freerun; Refer to Section~\ref{sec:LOS_experiment} for more explanation. Observe large ETx energy consumption gain for Beaconing and Probing, compared to Freerun.}
\label{fig:experiment_results_1}
\end{figure}

\subsection{Experimental Results: Case 0---Benchmark}
\label{sec:benchmark_measurements}

To obtain the metrics of interest from the measurements for the benchmark (Freerun protocol), see Section~\ref{sec:charger_energy_saving_protocol}, we use the measured values in $\mathbf{R_c}$, described in Section~\ref{sec:charge_accuracy_metric}, to calculate the harvested power at each position depicted in Fig.~\ref{fig:experiment_setup}. We then sum up the harvested power of four ETx, as the theoretical harvested power in the testing scenario where four ETxs are switched on all the time. Then we measure the same performance parameters as for the other two protocols. Results are presented in Fig.~\ref{fig:experiment_results_1} and Fig.~\ref{fig:experiment_results_2} and discussed in the subsequent sections. Note that all experimental results were plotted using MATLAB's \textsf{boxplot} function.

\subsection{Experimental Results: Case 1---Line of Sight Scenario}
\label{sec:LOS_experiment}

We have performed the experiment for five different communication threshold values, \txRssiThreshold, to measure WPTN performance simulating various ETx/ERx link qualities. The result is presented in Fig.~\ref{fig:experiment_results_1}.

\subsubsection{ERx Harvested Energy}

Refer to Fig.~\ref{fig:exp1_1}. For every value of \txRssiThreshold, the energy harvested by Beaconing is higher than for Probing. This is due to restriction of Probing, where at most one ETx can charge the ERx during a beacon period. The Beaconing protocol allows multiple ETx to be turned on at the same time.

As the \txRssiThreshold increases, the harvested energy decreases for both protocols---Beaconing and Probing. Naturally, the higher the threshold is, the less probability that the ETx would be triggered on by neighboring ERx. As expected, the Freerun mode has the highest harvested energy in almost every testing point, because all ETxs are switched on all the time.

\subsubsection{ETx Energy Consumption}

We are now ready to present the fundamental result of this paper, \emph{proving the ``green'' aspect of the designed WPTN}. In addition to harvested energy we show total energy used (in kJ) by all ETx during the whole experiment, refer to Fig.~\ref{fig:exp1_0}. We clearly see the power saved by the Beaconing and Probing protocol, compared with the Freerun mode (for Probing---by almost five times). Since Freerun mode switches ETx all the time, the energy consumption by ETx is highest and constant over \txRssiThreshold. We discuss the reason behind this gain in detail in subsequent sections.

\subsubsection{ERx Energy Consumption}

The power consumption of the Probing protocol is higher than for Beacon in our measurements. The main reason is that Probing needs ERx to receive the probing command from the ETx, measure the signal strength and send feedback packets to the charger. The ETxs request ERx to measure harvested power in every beacon round. Note that Probing protocol uses three message types to trigger the charging Phase, see Protocol~\ref{alg:erx_states}, while Beaconing protocol uses only one message to trigger charging, see Protocol~\ref{alg:beaconing_implementation}.

As the \txRssiThreshold increases, the power consumption by the ERx with Probing decreases. The reason is that the larger the \txRssiThreshold is, the less probability that ETx will accept charge request messages from the ERx. Then larger \txRssiThreshold causes less number of ETx to associate with the ERx on probing, which further causes less number of communication messages at ERx.

\subsubsection{WPTN Efficiency}

Compared with the Beaconing protocol, Probing stays on a stable level for each \txRssiThreshold. At --70\,dBm, the efficiency of Probing is around three times larger than Beaconing. At --50\,dBm, the efficiency of Beaconing increases. The main reason is that \txRssiThreshold~represents the range that ETx evaluates whether the ERx can be successfully charged or not. The larger the \txRssiThreshold is, the smaller the threshold range is, and the higher energy ERx can harvest in the range. Therefore, smaller \txRssiThreshold value results in higher efficiency. The benefit of using smaller threshold \txRssiThreshold is that the power transmission efficiency increases. The drawback is that decreasing range causes decreasing harvested energy, see Fig~\ref{fig:exp1_1}.

The Freerun mode always takes the lowest charging efficiency because it can not estimate whether ERx is inside or outside the WPTN and what is the possible harvesting power. If the receiver is outside the WPTN or in the area with very low power radio, switching on ETxs will waste a lot of power. In the experiment, the disappearing time of ERx is 15\,s. We conjecture that if the disappearing time increases the efficiency of Freerun mode will be even lower.


\subsubsection{WPTN Accuracy}
\label{sec:accuracy_los}

Probing protocol keeps relatively high and stable accuracy from --70\,dBm to --50\,dBm. High accuracy causes the high efficiency in probing based protocol as shown in Fig~\ref{fig:exp1_2}. In Probing, only one ETx is allowed to charge the ERx which potentially decreases the accuracy. We hypothesize that if multiple ETxs could exploit Probing-like protocol at the same time the accuracy and efficiency could further increase.

The accuracy of Beaconing increases as the threshold increases from --70\,dBm to --50\,dBm---the higher \txRssiThreshold value is, the closer the ERx must be to an ETx for triggering the charging. And the closer the ERx is to the charger, the higher the probability that the ETx can charge the ERx. Freerun mode naturally has the worst charging accuracy---the ERx can hardly harvest sufficient energy at certain positions while ETxs are continuously switched on.

\subsubsection{ERx Time to Charge}

\begin{figure}
\centering
\subfigure[Beaconing, $N=4$, $K=1$]{\includegraphics[width=0.24\textwidth]{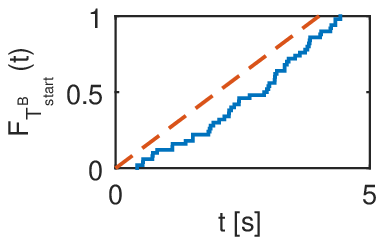}\label{fig:beaconing_time_n4k1}}
\subfigure[Beaconing, $N=4$, $K=2$]{\includegraphics[width=0.24\textwidth]{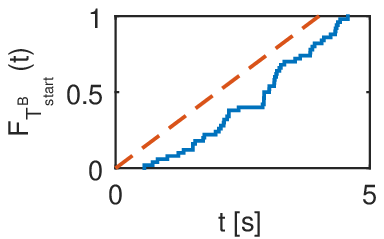}\label{fig:beaconing_time_n4k2}}
\subfigure[Beaconing, $N=4$, $K=3$]{\includegraphics[width=0.24\textwidth]{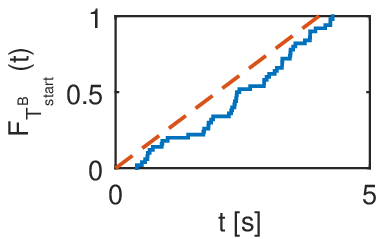}\label{fig:beaconing_time_n4k3}}
\subfigure[Beaconing, , $N=4$, $K=4$]{\includegraphics[width=0.24\textwidth]{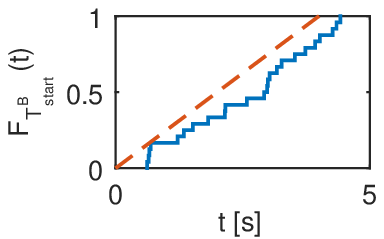}\label{fig:beaconing_time_n4k4}}
\subfigure[Probing, $N=4$, $K=1$]{\includegraphics[width=0.24\textwidth]{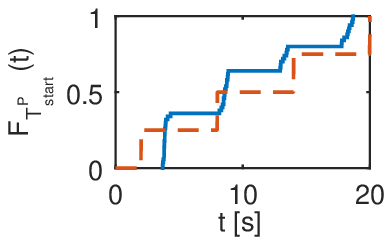}\label{fig:probing_time_n4k1}}
\subfigure[Probing, $N=4$, $K=2$]{\includegraphics[width=0.24\textwidth]{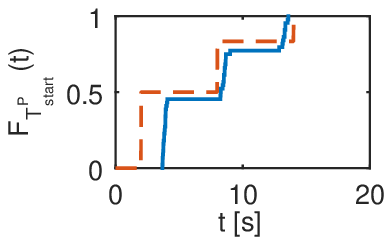}\label{fig:probing_time_n4k2}}
\subfigure[Probing, $N=4$, $K=3$]{\includegraphics[width=0.24\textwidth]{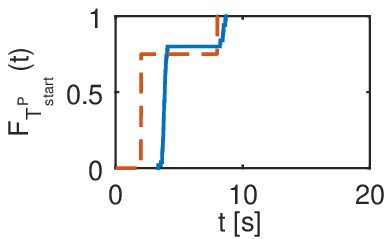}\label{fig:probing_time_n4k3}}
\subfigure[Probing, $N=4$, $K=4$]{\includegraphics[width=0.24\textwidth]{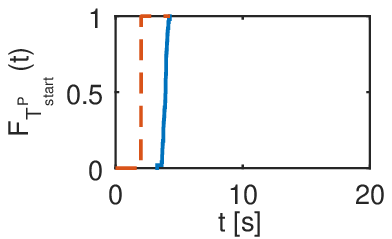}\label{fig:probing_time_n4k4}}
\caption{WTPN time to charge CDF: P---Probing, B---Beaconing, solid line--experiment, dashed line--analysis.}
\label{fig:time_to_charge_results}
\end{figure}

\begin{figure}
\centering
\subfigure[ERx harvested energy]{\includegraphics[width=0.33\columnwidth]{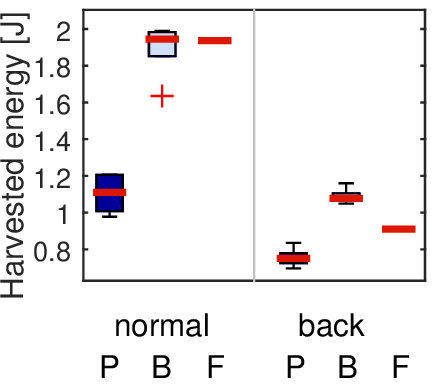}\label{fig:exp2_1}}
\subfigure[ETx consumed energy]{\includegraphics[width=0.33\columnwidth]{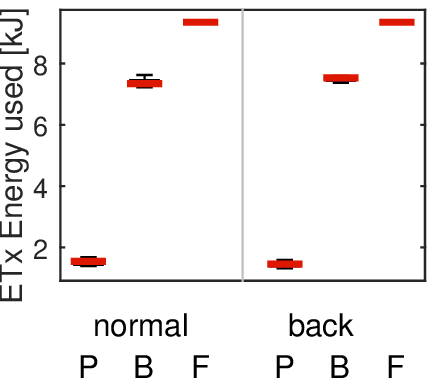}\label{fig:exp2_0}}
\subfigure[WPTN Efficiency]{\includegraphics[width=0.33\columnwidth]{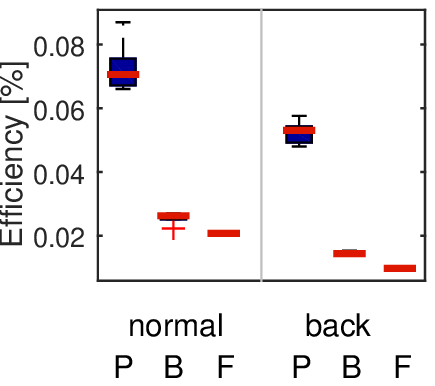}\label{fig:exp2_2}}
\subfigure[ERx Communication cost]{\includegraphics[width=0.33\columnwidth]{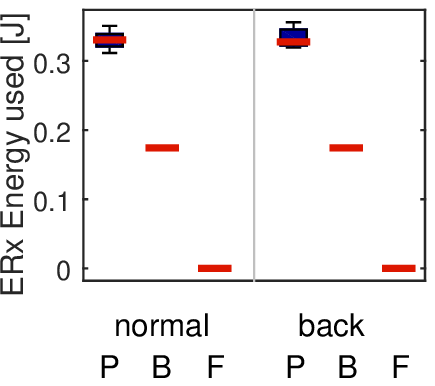}\label{fig:exp2_3}}
\subfigure[WPTN Accuracy]{\includegraphics[width=0.33\columnwidth]{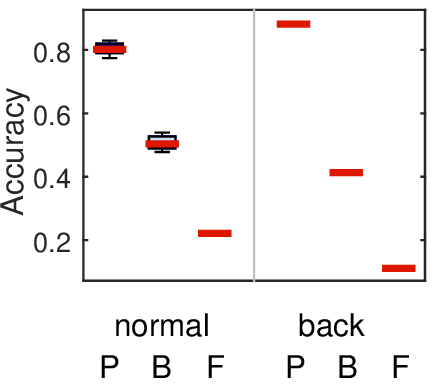}\label{fig:exp2_4}}
\caption{WTPN experiment results: (a); B---Beaconing, P---Probing, F---Freerun; Refer to Section~\ref{sec:non-LOS_experiment} for more explanation.}
\label{fig:experiment_results_2}
\end{figure}

To verify theoretical analysis of time to charge in both protocols, we have conducted an experiment, where we have placed ERx less than 50\,cm to each of ETx devices (to ensure ERx is within charging range of all ETxs). To emulate ERx being within charging range of a given ERx, we would connect or disconnect Powercast device from Arduino microcontroller. For each value of $K$ from $K=1$ (one ETx connected) to $K=4$ (four ETxs connected) we have performed an experiment where ERx appears randomly in the network 50 times. Afterwards, we measured the time it takes from appearing in the network to being charged. A CDF values of those experiments are presented in Fig.~\ref{fig:time_to_charge_results}. In this figure experimental results (solid lines) are compared against theoretical results (dashed lines). For Fig.~\ref{fig:beaconing_time_n4k1}--Fig.~\ref{fig:beaconing_time_n4k4} additionally a CDF of $T^{\text{B}}_{\text{start}}$ is added.

We see that Beaconing is faster in reaching ETx than Probing, however with increasing $K$ time to charge for Probing becomes very low as well (almost instant connection after approximately two seconds). For Beaconing, irrespective of number of ETx, the time to charge stays constant. The discrepancy between experimental and numerical results is due to approximation of not taking into account propagation and processing time. Nevertheless the analysis follow the trends of the experimental results in all cases reasonably well.

\subsection{Experimental Results: Case 2---Non-Line of Sight Scenario}
\label{sec:non-LOS_experiment}

In this experiment, we have tested the performance of WPTN, in which the ERx is inside the communication range while outside the charging range of ETx. We change the experiment setup in Fig.~\ref{fig:wptn_setup} by turning ETx 1 and 3 by 180 degrees around its axis. Results in this testing scenario and previous section are depicted as \textit{back}, and \textit{normal}. ETx \txRssiThreshold is set to --70\,dBm.

\subsubsection{ERx Harvested Energy}
\label{sec:harvested_energy_nlos}

In both normal and back condition, the harvested energy of Beaconing protocol is larger than the Probing protocol. By Beaconing protocol the harvested energy in back condition decreases by 45\% from the normal condition, which fits with the experiment setup by turning two chargers 180 degrees back. The decreasing percent in Probing from normal to back condition is 32\%, which is smaller than in Beaconing protocol. This is because Probing selects the ETx that can charge energy over the threshold \txRssiThreshold.

\subsubsection{ETx Energy Consumption}
\label{sec:etx_energy_consumption}

As expected, the power consumption of ETxs in Beacon protocol and Freerun mode maintains at the same level in both normal and back conditions. Beacon protocol does not give ETxs the function to know whether the charging power is efficiently harvested not. Therefore, as in the back condition, even if the ERx is outside the charging range, the charger still switches on as it hears the request message from the ERx. Lots of energy is wasted in the back condition by Beaconing protocol. In Probing protocol, the `back' ETxs (ETx 1 and ETx 3) will switch off, after they evaluate the harvested energy at the ERx is low.

\subsubsection{ERx Energy Consumption}
\label{sec:power_consumption_nlos}

For both protocols, Beaconing and Probing, the scheduling of the communication in ERx are not influenced by the topology of WPTN. So the power consumption in communication in both normal and back conditions are almost the same.

\subsubsection{WPTN Efficiency}
\label{sec:efficiency_nlos}

By Beaconing protocol, the efficiency in the back condition decreases by 45\% from the normal condition.
The decrease in Probing protocol from normal to back condition is 25\%, which is much smaller than for Beaconing. The smaller decreasing percent is because Probing protocol makes ETx work only when the charged energy is over \txRssiThreshold. The 25\% decrease mainly comes from the probing period when the ETx is turned on and asks the ERx to measure the harvested power. In our hardware implementation the probing period is very short (4\,s) and we speculate that increasing the probing period can further increase the efficiency of probing based protocol in the back condition.

\subsubsection{WPTN Accuracy}
\label{sec:accuracy_nlos}

The accuracy trends in the non-line of sight scenario are the same as in line-of-sight case, Sec.~\ref{sec:accuracy_los}. Also, as in the previous case Probing protocol obtains the highest accuracy and confirms its supremacy over the other two considered approaches.

\section{Discussion: Limitations and Further Research Directions}
\label{sec:discussion}

With these results we do believe we open up a new research direction within WPTN and we are aware of many points of improvement. We list the most important ones here.

\begin{enumerate}
\item Considering Beaconing protocol, the charge request rate should be optimized with respect to power consumption and harvested energy of the ERx. For example, the beacon period can adapt to the number of ERxs in the WPTN. Since as long as one ERx calls the charger to switch on there is no need to trigger ETXs for every ERx.

\item Considering Probing protocol, the probing frequency should be optimized with respect to power consumption of probing as well. ERxs should optimize the probing scheduling considering static and dynamic conditions. For example, when the ERx is static, there is no need to make the receiver measure harvested power with high frequency, since the harvested power is not expected to fluctuate much in such case.

\item WPTN should optimize the combinations of the subset of switched on ETxs to take advantage of the constructive signal combined at the ERx. Measuring all possible combination of the subset of the neighboring ETxs to switch on consumes too much time and power at the ERx. Thus a novel charge control algorithms are required to enable green WPTN with multiple ETxs operating at a time.

\end{enumerate}

\section{Conclusions}
\label{sec:conclusions}

In this paper we have introduced a new class of charging control protocol for wireless power transfer networks (WPTNs)---denoted as `green'---that conserve energy of the chargers. The purpose of such protocol is to maximize three metrics of interest to WPTNs (that we introduced here for the first time): (i) ETx charge accuracy, (ii) ETx charge efficiency and (iii) ERx harvested power, which in-turn minimize unnecessary uptime of WPTN energy transmitters. We prove that this problem is NP-hard. 

To solve it we propose two heuristics, denoted as `beaconing' (where energy receivers simply request power from transmitters) and `probing' (based on the principle of charge feedback from the energy receivers to the energy transmitters). The strength of our protocols lies in making few assumptions about the WTPN environment.

We conclude that each protocol performs its task best in two special cases. Experimentally we show that for large distances between chargers and receivers, probing is more efficient and accurate but harvests less energy than beaconing and has higher communication cost. As the charger to receiver distance increases, the efficiency of the beaconing-based protocol increases (since communication range is well correlated with charging range).

\section*{Acknowledgments}
\label{sec:acknowledgment}

We would like to thank Dr. Roberto Riggio from Create-Net for providing the Energino platform, Prof. Joshua Smith's group at the University of Washington for providing Impinj R1000 RFID reader, and to Mark Stoopman from TU Delft for help at the initial stage of the project.


\begin{thebibliography}{10}
\providecommand{\url}[1]{#1}
\csname url@samestyle\endcsname
\providecommand{\newblock}{\relax}
\providecommand{\bibinfo}[2]{#2}
\providecommand{\BIBentrySTDinterwordspacing}{\spaceskip=0pt\relax}
\providecommand{\BIBentryALTinterwordstretchfactor}{4}
\providecommand{\BIBentryALTinterwordspacing}{\spaceskip=\fontdimen2\font plus
\BIBentryALTinterwordstretchfactor\fontdimen3\font minus
  \fontdimen4\font\relax}
\providecommand{\BIBforeignlanguage}[2]{{%
\expandafter\ifx\csname l@#1\endcsname\relax
\else
\language=\csname l@#1\endcsname
\fi
#2}}
\providecommand{\BIBdecl}{\relax}
\BIBdecl

\bibitem{cherry_spectrum_2004}
S.~{Cherry}, ``Edholm's law of bandwidth,'' \emph{{IEEE} Spectr.}, vol.~41,
  no.~7, pp. 58--60, Jul. 2004.

\bibitem{witricity_website}
\BIBentryALTinterwordspacing
(2014) WiTricity Homepage. [Online]. Available: \url{http://www.witricity.com}
\BIBentrySTDinterwordspacing

\bibitem{ubeam_website}
\BIBentryALTinterwordspacing
(2014) UBeam Homepage. [Online]. Available: \url{http://www.ubeam.com}
\BIBentrySTDinterwordspacing

\bibitem{ossia_website}
\BIBentryALTinterwordspacing
(2014) Ossia Homepage. [Online]. Available: \url{http://www.ossiainc.com}
\BIBentrySTDinterwordspacing

\bibitem{artemis_website}
\BIBentryALTinterwordspacing
(2014) Artemis Home Page. [Online]. Available: \url{http://www.artemis.com}
\BIBentrySTDinterwordspacing

\bibitem{energous_website}
\BIBentryALTinterwordspacing
(2014) Energous Homepage. [Online]. Available: \url{http://energous.com}
\BIBentrySTDinterwordspacing

\bibitem{proxi_website}
\BIBentryALTinterwordspacing
(2014) Homepage. [Online]. Available: \url{http://powerbyproxi.com}
\BIBentrySTDinterwordspacing

\bibitem{dai_tpds_2014}
\BIBentryALTinterwordspacing
H.~{Dai}, G.~{Chen}, C.~{Wang}, S.~{Wang}, X.~{Wu}, and F.~{Wu}, ``Quality of
  energy provisioning for wireless power transfer,'' \emph{{IEEE} Trans.
  Parallel Distrib. Syst.}, 2014, accepted for publications. [Online].
  Available:
  \url{http://ieeexplore.ieee.org/stamp/stamp.jsp?tp=&arnumber=6762897}
\BIBentrySTDinterwordspacing

\bibitem{xie_wcom_2013}
L.~{Xie}, Y.~{Shi}, Y.~T. {Hou}, and W.~{Lou}, ``Wireless power transfer and
  applications to sensor networks,'' \emph{{IEEE} Wireless Commun. Mag.},
  vol.~20, no.~3, pp. 140--145, Aug. 2013.

\bibitem{wicaksono_vtc_2011}
R.~P. {Wicaksono}, G.~K. {Tran}, K.~{Sakaguchi}, and K.~{Araki}, ``Wireless
  grid: Enabling ubiquitous sensor networks with wireless energy supply,'' in
  \emph{Proc. IEEE VTC-Spring}, Yokohama, Japan, May 15--18, 2011.

\bibitem{he_tmc_2013}
S.~{He}, J.~{Chen}, F.~{Jiang}, D.~K. {Yau}, G.~{Xing}, and Y.~{Sun}, ``Energy
  provisioning in wireless rechargeable sensor networks,'' \emph{{IEEE} Trans.
  Mobile Comput.}, vol.~12, no.~10, pp. 1931--1942, Oct. 2013.

\bibitem{thomas_jbcs_2012}
S.~J. {Thomas}, R.~R. {Harrison}, A.~{Leonardo}, and M.~S. {Reynolds}, ``A
  battery-free multichannel digital {Neural/EMG} telemetry system for flying
  insects,'' \emph{{IEEE} Trans. Biomed. Circuits Syst.}, vol.~6, no.~5, pp.
  424--435, Oct. 2012.

\bibitem{greene_unpublished_2010}
\BIBentryALTinterwordspacing
C.~G. D. H.~D. {Kalp} and W.~{Tauche}, ``Making wireless sensor networks truly
  wireless using {RF} power,'' 2010. [Online]. Available:
  \url{http://www.sensormgmt.com/Articles/Powered%20By%20FireFly60614.pdf}
\BIBentrySTDinterwordspacing

\bibitem{holleman_biocas_2008}
J.~{Holleman}, D.~{Yeager}, R.~{Prasad}, J.~R. {Smith}, and B.~{Otis},
  ``{NeuralWISP}: An energy-harvesting wireless neural interface with 1-m
  range,'' in \emph{Proc. IEEE BioCAS}, Baltimore, MD, USA, Nov. 20--22, 2008.

\bibitem{long_cicc_2008}
J.~R. {Long}, W.~{Wu}, Y.~{Dong}, Y.~{Zhao}, M.~A.~T. {Sanduleanu},
  J.~{Gerrits}, and G.~van {Veenendaal}, ``Energy-efficient wireless front-end
  concepts for ultra lower power radio,'' in \emph{Proc. IEEE CICC}, San Jose,
  CA, USA, Sep. 21--24, 2008.

\bibitem{patel_wcm_2010}
M.~{Patel} and J.~{Wang}, ``Applications, challenges, and prospective in
  emerging body area networking technologies,'' \emph{{IEEE} Wireless Commun.
  Mag.}, vol.~17, no.~1, pp. 80--88, Feb. 2010.

\bibitem{michael_science_2011}
\BIBentryALTinterwordspacing
J.-B. {Michel}, Y.~K. {Shen}, A.~P. {Aiden}, A.~{Veres}, M.~K. {Gray},
  W.~{Brockman}, {The Google Books Team}, J.~P. {Pickett}, D.~{Hoiberg},
  D.~{Clancy}and Peter~{Norvig}, J.~{Orwant}, S.~{Pinker}, M.~A. {Nowak}, and
  E.~L. {Aiden}, ``Quantitative analysis of culture using millions of digitized
  books,'' \emph{Science}, vol. 331, no. 6014, pp. 176--182, Jan. 2011.
  [Online]. Available:
  \url{http://www.sciencemag.org/content/331/6014/176.full.pdf}
\BIBentrySTDinterwordspacing

\bibitem{massa_procieee_2013}
A.~{Massa}, G.~{Oliveri}, F.~{Viani}, and P.~{Rocca}, ``Array designs for
  long-distance wireless power transmission: State-of-the-art and innovative
  solutions,'' \emph{Proc. {IEEE}}, vol. 101, no.~6, pp. 1464--1481, Jun. 2013.

\bibitem{Pinuela_mtt_2013}
M.~{Pi{\~{n}}uela}, P.~D. {Mitcheson}, and S.~{Lucyszyn}, ``Ambient {RF} energy
  harvesting in urban and semi-urban environments,'' \emph{{IEEE} Trans.
  Microwave Theory Tech.}, vol.~61, no.~7, pp. 2715--2726, Jul. 2013.

\bibitem{timotheou_twc_2014}
S.~{Timotheou}, I.~{Krikidis}, G.~{Zheng}, and B.~{Ottersten}, ``Beamforming
  with {MISO} interference channels with {QoS} and {RF} energy transfer,''
  \emph{{IEEE} Trans. Wireless Commun.}, vol.~13, no.~5, pp. 2646--2658, May
  2014.

\bibitem{r1000_data_sheet}
\BIBentryALTinterwordspacing
{Impinj Inc.} (2015) Impinj speedway {R1000} {RFID} reader data sheet.
  [Online]. Available: \url{http://www.mas-rfid-solutions.com/impinj.pdf}
\BIBentrySTDinterwordspacing

\bibitem{s420_data_sheet}
\BIBentryALTinterwordspacing
------. (2015) Impinj speedway r1000 rfid reader data sheet. [Online].
  Available:
  \url{https://support.impinj.com/hc/en-us/articles/202755298-Reader-Documentation}
\BIBentrySTDinterwordspacing

\bibitem{powercast_website}
\BIBentryALTinterwordspacing
(2014) Powercast Corp. Homepage. [Online]. Available:
  \url{http://www.powercastco.com}
\BIBentrySTDinterwordspacing

\bibitem{ti_website}
\BIBentryALTinterwordspacing
(2014) Texas Instruments Homepage. [Online]. Available: \url{http://www.ti.com}
\BIBentrySTDinterwordspacing

\bibitem{sllrp_github}
\BIBentryALTinterwordspacing
B.~Ransford. (2015) Llrp library controller. [Online]. Available:
  \url{https://github.com/ransford/sllurp}
\BIBentrySTDinterwordspacing

\bibitem{lu_arxiv_2014}
\BIBentryALTinterwordspacing
X.~{Lu}, P.~{Wang}, D.~{Niyato}, D.~I. {Kim}, and Z.~{Han}, ``Wireless networks
  with {RF} energy harvesting: A contemporary survey,'' Jun. 26, 2014.
  [Online]. Available: \url{http://arxiv.org/abs/1406.6470}
\BIBentrySTDinterwordspacing

\bibitem{gomez_wiopt_2012}
\BIBentryALTinterwordspacing
K.~{Gomez}, R.~{Riggio}, T.~{Rasheed}, D.~{Miorandi}, and F.~{Granelli},
  ``Energino: a hardware and software solution for energy consumption
  monitoring,'' in \emph{Proc. International Workshop on Wireless Network
  Measurements}, Paderborn, Germany, May 18, 2012. [Online]. Available:
  \url{http://www.energino-project.org}
\BIBentrySTDinterwordspacing

\bibitem{gollakota_computer_2014}
S.~{Gollakota}, M.~S. {Reynolds}, J.~R. {Smith}, and D.~J. {Wetherall}, ``The
  emergence of {RF}-powered computing,'' \emph{{IEEE} Computer}, vol.~47,
  no.~1, pp. 32--39, Jan. 2014.

\bibitem{huang_arxiv_2012}
\BIBentryALTinterwordspacing
K.~{Huang} and V.~K.~N. {Lau}, ``Enabling wireless power transfer in cellular
  networks: Architecture, modeling and deployment,'' Jul. 24, 2012, submitted
  to {IEEE} Trans. Wireless Commun. [Online]. Available:
  \url{http://arxiv.org/abs/1207.5640}
\BIBentrySTDinterwordspacing

\bibitem{ju_twc_2014}
H.~{Ju} and R.~{Zhang}, ``Throughput maximization for wireless powered
  communication networks,'' \emph{{IEEE} Trans. Wireless Commun.}, vol.~13,
  no.~1, pp. 418--428, Jan. 2014.

\bibitem{krikidis_tcom_2013}
I.~{Krikidis}, ``Simultaneous information and energy transfer in large-scale
  networks with/without relaying,'' \emph{{IEEE} Trans. Commun.}, vol.~62,
  no.~3, pp. 900--912, Mar. 2014.

\bibitem{yang_arxiv_2013}
\BIBentryALTinterwordspacing
G.~{Yang}, C.~K. {Ho}, and Y.~L. {Guan}, ``Dynamic resource allocation for
  multiple-antenna wireless power transfer,'' Nov. 17, 2013, submitted to
  {IEEE} Trans. Signal Processing. [Online]. Available:
  \url{http://arxiv.org/abs/1311.4111}
\BIBentrySTDinterwordspacing

\bibitem{ng_globecom_2014}
\BIBentryALTinterwordspacing
D.~W.~K. {Ng} and R.~{Schober}, ``Resource allocation for coordinated
  multipoint networks with wireless information and power transfer,'' in
  \emph{Proc. IEEE GLOBECOM}, Austin, TX, USA, Dec. 8--12, 2014. [Online].
  Available: \url{http://arxiv.org/abs/1403.5730}
\BIBentrySTDinterwordspacing

\bibitem{fu_infocom_2013}
L.~{Fu}, P.~{Cheng}, Y.~{Gu}, J.~{Chen}, and T.~{He}, ``Minimizing charging
  delay in wireless rechargeable sensor network,'' in \emph{Proc. IEEE
  INFOCOM}, Turin, Italy, Apr. 14--19, 2013.

\bibitem{liu_net_2014}
\BIBentryALTinterwordspacing
X.~{Liu}, P.~{Wang}, D.~{Niyato}, and Z.~{Han}, ``Resource allocation in
  wireless networks with {RF} energy harvesting and transfer,'' \emph{{IEEE}
  Network}, May 22, 2014, accepted for publication. [Online]. Available:
  \url{http://arxiv.org/abs/1405.5630}
\BIBentrySTDinterwordspacing

\bibitem{mercier_jssc_2011}
P.~P. {Mercier} and A.~P. Chandrakasan, ``A supply-rail-coupled {eTextiles}
  transciever for body-area networks,'' \emph{{IEEE} J. Solid-State Circuits},
  vol.~46, no.~6, pp. 1284--1295, Jun. 2011.

\bibitem{wang_tmc_2014}
C.~{Wang}, J.~{Li}, and Y.~{Yang}, ``{NETWRAP}: An {NDN} based real-time
  wireless recharging framework for wireless sensor network,'' \emph{{IEEE}
  Trans. Mobile Comput.}, vol.~13, no.~6, pp. 1283--1297, Jun. 2014.

\bibitem{naderi_twc_2014}
M.~Y. {Naderi}, P.~{Nintanavongsa}, and K.~R. {Chowdhury}, ``{RF-MAC}: A medium
  access control protocol for re-chargeable sensor networks powered by wireless
  energy harvesting,'' \emph{{IEEE} Trans. Wireless Commun.}, vol.~13, no.~7,
  pp. 3926--3937, Jul. 2014.

\bibitem{xiang_pimrc_2013}
L.~{Xiang}, J.~L.~K. {Han}, and G.~{Shi}, ``Fueling wireless networks
  perpetually: A case of multi-hop wireless power distribution,'' in
  \emph{Proc. IEEE PIRMC}, London, UK, Sep. 8--11, 2013.

\bibitem{yoon_ccnc_2013}
S.~K. {Yoon}, S.~J. {Kim}, and U.~K. {Kwon}, ``Energy relaying in mobile
  wireless sensor networks,'' in \emph{Proc. IEEE CCNC}, Las Vegas, NV, USA,
  Jan. 11--14, 2013.

\bibitem{dai_infocom_2014}
H.~{Dai}, Y.~{Liu}, G.~{Chen}, X.~{Wu}, and T.~{He}, ``Safe charging for
  wireless power transfer,'' in \emph{Proc. IEEE INFOCOM}, Toronto, Canada,
  Apr. 27~--~May 2, 2014.

\bibitem{visser_procieee_2013}
H.~J. {Visser} and R.~J.~M. {Vullers}, ``{RF} energy harvesting and transport
  for wireless sensor network applications: Principles and requirements,''
  \emph{Proc. {IEEE}}, vol. 101, no.~6, pp. 1410--1423, Jun. 2013.

\bibitem{strassner_procieee_2013}
B.~{Strassner, II} and K.~{Chang}, ``Microwave power transmission: Historical
  milestones and system components,'' \emph{Proc. {IEEE}}, vol. 101, no.~6, pp.
  1379--1396, Jun. 2013.

\bibitem{sample_procieee_2013}
A.~P. {Sample}, B.~H. {Waters}, S.~T. {Wisdom}, and J.~R. {Smith}, ``Enabling
  seamless wireless power delivery in dynamic environments,'' \emph{Proc.
  {IEEE}}, vol. 101, no.~6, pp. 1343--1358, Jun. 2013.

\bibitem{ho_procieee_2013}
J.~S. {Ho}, S.~{Kim}, and A.~S.~Y. {Poon}, ``Midfield wireless powering for
  implantable systems,'' \emph{Proc. {IEEE}}, vol. 101, no.~6, pp. 1369--1378,
  Jun. 2013.

\bibitem{lu:2014:arxiv}
\BIBentryALTinterwordspacing
X.~{Lu}, D.~{Niyato}, P.~{Wang}, D.~I. {Kim}, and Z.~{Han}, ``Wireless charger
  networking for mobile devices: Fundamentals, standards, and applications,''
  Oct. 31, 2014. [Online]. Available: \url{http://arxiv.org/abs/1410.6635}
\BIBentrySTDinterwordspacing

\bibitem{zhao_tmc_2014}
M.~{Zhao}, J.~{Li}, and Y.~{Yang}, ``A framework of joint mobile energy
  replenishment and data gathering in wireless rechargeable sensor networks,''
  \emph{{IEEE} Trans. Mobile Comput.}, vol.~3, no.~12, pp. 2689--2705, Dec.
  2014.

\bibitem{liu_tcom_2013}
L.~{Liu}, R.~{Zhang}, and K.-C. {Chua}, ``Wireless information and power
  transfer: A dynamic power splitting approach,'' \emph{{IEEE} Trans. Commun.},
  vol.~61, no.~9, pp. 3990--4001, Sep. 2013.

\bibitem{bi:2014:arxiv}
\BIBentryALTinterwordspacing
S.~{Bi}, C.~K. {Ho}, and R.~{Zhang}, ``Wireless powered communication:
  Opportunities and challenges,'' Dec. 29, 2014, accepted for publication.
  [Online]. Available: \url{http://arxiv.org/abs/1408.2335}
\BIBentrySTDinterwordspacing

\bibitem{chiu_apnoms_2012}
T.-C. {Chiu}, Y.-Y. {Shih}, A.-C. {Pang}, J.-Y. {Jeng}, and P.-C. {Hsiu},
  ``Mobility-aware charger deployment for wireless rechargeable sensor
  network,'' in \emph{Proc. APNOMS}, Seoul, South Korea, Sep. 15--27, 2012.

\bibitem{xbee_website}
\BIBentryALTinterwordspacing
(2014) Digi International Homepage. [Online]. Available:
  \url{http://www.digi.com/xbee/}
\BIBentrySTDinterwordspacing

\bibitem{wisp_website}
\BIBentryALTinterwordspacing
(2014) WISP 5.0 Wiki. [Online]. Available: \url{http://wisp5.wikispaces.com}
\BIBentrySTDinterwordspacing

\bibitem{Garey:1979:bell}
M.~R. {Garey} and D.~S. {Johnson}, \emph{Computers and Intractability: A Guide
  to the Theory of {NP}-Completeness}.\hskip 1em plus 0.5em minus 0.4em\relax
  New York, NY, USA: Bell Telephone Labolatories, 1979.

\bibitem{kleinberg:2005:algo_design}
J.~{Kleinberg} and {\'{E}}.~{Tardos}, \emph{Algorithm Design}.\hskip 1em plus
  0.5em minus 0.4em\relax Addison Wesley, 2005.

\bibitem{leeuwen:1990:handbook}
D.~S. {Johnson}, ``Handbook of theoretical computer science: Algorithms and
  complexity,'' J.~van {Leeuwen}, Ed.\hskip 1em plus 0.5em minus 0.4em\relax
  Elsevier, 1990, vol.~A.

\bibitem{freville2004multidimensional}
A.~Fr{\'e}ville, ``The multidimensional 0-1 knapsack problem: An overview,''
  \emph{European Journal of Operational Research}, vol. 155, no.~1, pp. 1--21,
  2004.

\bibitem{arduino_website}
\BIBentryALTinterwordspacing
(2014) Arduino Homepage. [Online]. Available: \url{http://arduino.cc/en/Main/}
\BIBentrySTDinterwordspacing

\bibitem{golinski_msc_2015}
M.~Goli{\'n}ski, ``Wireless power transfer networks,'' Master's thesis, Delft
  University of Technology, Delft, the Netherlands, 2015.

\bibitem{casilari_sensors_2010}
\BIBentryALTinterwordspacing
E.~{Casilari}, J.~M. {Cano-Garcia}, and G.~{Campos-Garrido}, ``Modeling of
  current consumption in {802.15.4/ZigBee} sensor motes,'' \emph{Sensors},
  vol.~10, no.~6, pp. 5443--5468, Jun. 2010. [Online]. Available:
  \url{http://www.mdpi.com/1424-8220/10/6/5443}
\BIBentrySTDinterwordspacing

\bibitem{avr_data_sheet}
\BIBentryALTinterwordspacing
{Atmel Corp.} (2015) Atmel 8-bit microcontroller with 4/8/16/32\,kbytes
  in-system programmable flash. [Online]. Available:
  \url{http://www.atmel.com/images/Atmel-8271-8-bit-AVR-Microcontroller-ATmega48A-48PA-88A-88PA-168A-168PA-328-328P_datasheet_Complete.pdf}
\BIBentrySTDinterwordspacing

\bibitem{xbee_data_sheet}
\BIBentryALTinterwordspacing
{Digi International Inc.}, ``Xbee 802.15.4 module specification,'' 2015.
  [Online]. Available:
  \url{http://www.digi.com/products/wireless-wired-embedded-solutions/zigbee-rf-modules/point-multipoint-rfmodules/xbee-series1-module#specs}
\BIBentrySTDinterwordspacing

\end{thebibliography}
\end{document}